\documentclass{emulateapj}
\usepackage{natbib}

\begin{document}

\title{The H$\alpha$-based Star Formation Rate Density of the Universe
  at z=0.84.}

\author{V\'{i}ctor Villar, Jes\'us Gallego, Pablo G. P\'erez-Gonz\'alez, Sergio Pascual}

\affil{Universidad Complutense de Madrid, E-28040 Madrid, Spain}
\email{viv,jgm,pgperez,spr@astrax.fis.ucm.es}

\author{Kai Noeske, David C. Koo} 

\affil{Lick Observatory, University of California, Santa Cruz, CA
95064}
\email{koo,kai@ucolick.org}
\and

\author{Guillermo Barro, Jaime Zamorano} 

\affil{Universidad Complutense de Madrid, E-28040 Madrid, Spain}
\email{gbc,jaz@astrax.fis.ucm.es}
\begin{abstract}
We present the results of an H$\alpha$ {\it near-infrared} narrow-band
survey searching for star-forming galaxies at redshift $z=0.84$. This
work is an extension of our previous narrow-band studies in the
optical at lower redshifts. After removal of stars and redshift
interlopers (using spectroscopic and photometric redshifts), we build
a complete sample of 165 H$\alpha$ emitters in the Extended Groth
strip and GOODS-N fields with
L(H$\alpha$)$>$10$^{41}$~erg~s$^{-1}$. We compute the H$\alpha$
luminosity function at $z=0.84$ after corrections for [NII] flux
contamination, extinction, systematic errors, and incompleteness. Our
sources present an average dust extinction of
A(H$\alpha$)=1.5~mag. Adopting H$\alpha$ as a surrogate for the
instantaneous star formation rate (SFR), we measure a
extinction-corrected SFR density of $0.17^{+0.03}_{-0.03}$ M$_{\odot}$
yr$^{-1}$ Mpc$^{-3}$. Combining this result to our prior measurements
at z$=$0.02, 0.24, and 0.40, we derive an H$\alpha$-based evolution of
the SFR density proportional to (1+z)$^\beta$ with $\beta=3.8\pm0.5$.
This evolution is consistent with that derived by other authors using
different SFR tracers.
\end{abstract}


\section{Introduction}
The cosmic Star Formation Rate (SFR) density evolution of the Universe
is an important constraint on galaxy formation and evolution
models. Deep redshift surveys have proved that star-formation activity
substantially increases with redshift from z$=$0 to $z\simeq1$
\citep[see][for a review]{fer00}. This behavior has been reproduced by
current galaxy evolution theories
\citep[see the review by][]{Bau06}.

Several tracers can be used to obtain SFRs at different redshifts:
ultraviolet (UV) continuum, nebular lines such as [O{\sc
ii}]$\lambda$3727 or H$\alpha$, total infrared (TIR), or radio
continuum luminosities. For a summary of SFR density measurements, see
\cite{Hop04} and \cite{Hop06}. In the redshift regime from z$\sim$1 to
$z=0$, the observational data in the Far-IR and UV is now much more
robust with the results from {\it Spitzer} \citep{PG05} and those from
GALEX \citep{Arn05,Schimi05} and the VVDS \citep{tresse2007}.

Focusing on the H$\alpha$ SFR tracer, the local SFR density was first
measured by \citet[][see also \citealt{PG03}]{Gallego95} using the UCM
Survey \citep{Zam94,Zam96}. Similar values at z$=$0 have also been
obtained more recently by the SDSS \citep{Brinchmann04} and SINGG
\citep{Hanish06} projects. At z=0.24, \cite{Tresse98} and
\cite{P01} obtained similar SFR densities for a sample of CFRS
galaxies and a sample selected using a narrow band technique like
ours. Recently, \cite{Shi07} also used the narrow band technique at
this redshift, reaching fainter luminosities. \cite{Jones01} used
their narrow band counts obtained with a tunable filter to study the
redshift range z$=$0.0$-$0.4. \cite{gla04} also used a tunable filter
to detect a total of eight emission-line galaxies in the Hubble Deep
Field, three of them being H$\alpha$ emitters at $z=0.40$. At
$z\sim$1, \cite{gla99} obtained a pioneering result from near-IR
spectroscopy of eight CFRS galaxies in the 0.79$<$z$<$1.1 redshift
range. Their results were completed by \cite{Tresse02}, who obtained
near-IR spectroscopy with VLT for 30 galaxies with redshifts
0.5$<$z$<$1.1. \cite{Do06} have recently obtained an average H$\alpha$
luminosity for 38 galaxies at 0.77$<$z$<$1 by stacking near-IR spectra
where the H$\alpha$ emission was not individually detected (for most
of the targets). Aperture and luminosity bias corrections are needed
to compare SFR densities from such slit spectroscopy studies with
other data. Slit-less spectroscopy from HST data for galaxies in the
0.7$<$z$<$1.9 redshift range allowed the analysis of \cite{Yan99} and
\cite{Hop00} for 33 and 37 emission-line galaxies, respectively. \cite{Ly07} used
several narrow band filters to study emission-line galaxies at
different redshifts and through different emission-lines.  Finally,
\cite{Red07} have estimated the SFRd at z$=$2--3 based on UV and FIR 
luminosity functions, also predicting the H$\alpha$ luminosity
function from them.

Significant discrepancies have been found when comparing the values
obtained from different studies and tracers, due to dust extinction,
metallicity and different spatial origins of the emission. The
H$\alpha$ luminosity is an excellent tracer of the SFR
\citep{Ken98,Charlot01}. It is essential when computing the SFR of a
galaxy from its optical spectrum \citep{Mous06}. The H$\alpha$
luminosity shares with the UV and TIR emissions the dependence to the
Initial Mass Function (IMF).  H$\alpha$-based SFRs are affected by
obscuration but are not very sensitive to metallicity. The TIR is not
affected by dust attenuation, but it may miss the unobscured star
formation which may be an important fraction of the total in certain
galaxies \citep{PG06}. There are also large uncertainties linked to
the estimation of the TIR emission (from 8 to 1000~$\mu$m) from
monochromatic measurements (e.g., the 24~$\mu$m flux). In addition,
other sources different from the recent star formation (old stellar
populations, AGNs) contribute to the dust heating in unknown
(difficult to quantify) amounts. The UV luminosity not only traces the
current SFR but also relatively old stellar populations \citep{Cal05}
and is heavily affected by obscuration.  As shown by \cite{Bell03},
obscuration corrected H$\alpha$ is consistent, within a factor of 2,
with the summed SFRs estimated using the UV and TIR
luminosities. Consequently, H$\alpha$ observations of galaxy samples
with UV and TIR data provides an invaluable tool to understand the
evolution of the SFR and the role of obscuration in the determination
of global SFR for galaxies.

Our group measured the SFR density locally \citep{Gallego95} using a
sample of H$\alpha$-selected galaxies from the objective-prism UCM
Survey \citep{Zam94,Zam96}. We then extended this measurement to
$z\simeq0.24$ \citep{P01,P05} and also $z\simeq0.4$ \citep{P05}, the
maximum redshift for which H$\alpha$ can be reached with CCDs. To
select the H$\alpha$ emitters, we successfully used our own optical
narrow band filters tuned to the wavelength of the redshifted
H$\alpha$ line. The goal of this paper is to extend our previous work
to z=0.84 using a narrow band filter centered at 1.20$\mu$m.

This paper is structured as follows. In Section~2, we present the
data, the observations, and reduction process. In Section~3, we
describe the different steps to select the final sample, including our
simulations to analyze the sample biases. In Section~4, we describe
the procedure to obtain the H$\alpha$ fluxes for each galaxy. In
Section~5, we present the H$\alpha$ luminosity functions (corrected
and uncorrected for extinction) and the star formation rate
density. Finally, we summarize our results and conclusions in
Section~6.

A concordance cosmology is assumed throughout this paper with
$H_{\rm0}=70$ km s$^{-1}$ Mpc$^{-1}$, $\Omega_M=0.3$,
$\Omega_\Lambda=0.7$ (Lahav \& Liddle 2006). With this cosmology,
1$\arcsec$ at z=0.84 corresponds to 7.63 kpc, the typical surveyed
volume for a 1\% narrow-band filter is $\sim$130000 Mpc$^{3}$
$^{\square^{-1}}$ and the Universe is 6.44 Gyr old.

\section{Data}

\subsection{Observations}

This work is based on deep near-infrared imaging obtained with a
broad- and a narrow-band filters. The narrow-band filter is the
J-continuum ($Jc$) centered at 1.20$\mu$m, corresponding to H$\alpha$
at z=0.84. The broad-band filter is used to determine an approximate
continuum level near the H$\alpha$ emission-line.

The survey was carried out with the near-infrared camera
OMEGA-2000\footnote{http://www.mpia-hd.mpg.de/IRCAM/O2000/index.html}
on the 3.5m telescope at the Calar Alto Observatory (Almer\'{\i}a,
Spain) with the $J$ and narrow-band filters. OMEGA-2000 is equipped
with a 2k$\times$2k Hawaii-2 detector with 18$\mu$m pixels
(0$\farcs$45 on the sky, 15$\arcmin \times$15$\arcmin$ field of view).
Three different 15$\arcmin \times$15$\arcmin$ pointings were obtained,
two in the Extended Groth Strip (EGS) and another one in the
GOODS-North field, in April 2005 and May 2006. The characteristics of
these 3 pointings are shown in Table~\ref{tab_fields}.

Each field was observed with a dithering pattern consisting of 20
different positions with typical relative offsets of
20$\arcsec$-30$\arcsec$. After that sequence, the telescope starts
another observation block at the initial position (slightly offseted
to remove artifacts). With this combination of patterns, the telescope
visits 400 different positions without repeating anyone. For the
$J$-band, we co-added 15 images of 10 seconds each, for a total of 150
seconds at each position. For the narrow-band filter ($Jc$), 5 images
of 30 seconds were coadded at each position, for a total of 150
seconds as well. The total average exposure time per pixel was
$\sim$7.2~ks in the J band and $\sim$18~ks in the $Jc$ filter.

\begin{deluxetable*}{cccccccccc}
\tabletypesize{\scriptsize}
\tablecaption{Observed fields}
\tablehead{
\colhead{Field} & \colhead{$\alpha$ (J2000)} & \colhead{$\delta$ (J2000)} & \colhead{area}& \colhead{t$_{exp}$ NB} & \colhead{t$_{exp}$ BB} & \colhead{FHWM($\arcsec$) NB} & \colhead{FWHM($\arcsec$) BB}& \colhead{m$_{lim}$ NB} & \colhead{m$_{lim}$ BB}\\
\colhead{(1)} & \colhead{(2)} & \colhead{(3)} & \colhead{(4)}& \colhead{(5)} & \colhead{(6)} & \colhead{(7)} & \colhead{(8)}& \colhead{(9)} & \colhead{(10)}
}
\startdata
Groth2  &14 17 31 &+52 28 11&0.0468 &17850&7200 &1.1&0.9 & 20.54 & 22.30\\
Groth3  &14 18 14 &+52 42 15&0.0648 &18000&7200 &0.9&1.1 & 20.99 & 22.43\\
GOODS-N &12 36 40 &+62 12 16&0.0622 &20300&9000 &1.0&0.9 & 20.83 & 21.89\\
\enddata
\tablecomments{(1) Field name (2) RA (J2000) (3) DEC (J2000) (4) Area (square degrees) (5) narrow-band exposure time (s) (6) broad-band exposure time (s) (7) narrow-band FWHM ($\arcsec$) (8) broad-band FWHM ($\arcsec$) (9) narrow-band limiting J$_{VEGA}$ magnitude (3$\sigma$) (10) broad-band limiting J$_{VEGA}$ magnitude (3$\sigma$)}
\label{tab_fields}
\end{deluxetable*}

\subsection{Reduction}
We used a combination of the IRAF package {\sc XDIMSUM} and our own
dedicated software for the reduction of the data. In a first
iteration, the dithered images were dark subtracted and combined
without shifting to produce a master flat-field. Pixels marked as
cosmetic defects were not used in this computation. After the flat
field correction, the sky was subtracted. In this first iteration, we
used the median value of each image as the sky value. At this point,
we checked the photometry and seeing for each individual image,
discarding those images that presented the worst seeing or with low
object signal due to the presence of clouds or low transparency. We
then combined the remaining images to produce a final mosaic. In the
combination, for each final pixel in the image, we rejected pixel
values from the individual images that exceeded by 3$\sigma$ the mean
signal. This allowed us to get rid of cosmetic defects and cosmic
rays. With this first final image, we produced an object mask by
detecting all the sources with Sextractor
\citep{bertin96}. In a second iteration, we repeated the same process
except that we changed the method to construct the flat-field and the
sky images. This time, for the flat-field construction we combined the
science frames rejecting, in addition to cosmetic defects, object
pixels. The sky subtraction was performed with {\sc XDIMSUM}, taking
previous and subsequent images to compute the sky for each individual
image. Object pixels and cosmetic defects were also excluded in the
sky construction. In a third iteration, we normalized the science
images dividing by the sky images before creating the flat-field. The
new flat-field is then not affected by the shape of the sky. The rest
of the process is the same as in the second iteration, obtaining the
final science image.

The observing runs were not fully photometric and we had to use bright
2MASS stars to do the photometric calibration. We introduced a
color-term because our $J$-filter was not exactly the same as the
2MASS $J$ filter. However, the color term was very small in most
cases, with ($J$-$J$$_{2MASS}<$0.08). We estimated that zeropoint
errors were lower than $\sim$0.15~mag. The narrow-band filter was
calibrated using the $J$-band as reference, assuming that the mean
color for the bright objects was zero.

\subsection{Additional data}

In order to estimate photometric redshifts and extinctions, we have
also used complementary datasets in both the GOODS-N and the EGS
fields. For the GOODS-N, we have used optical and NIR data spanning
from the U- to the $HK_{s}$-bands
\citep[UBVRIzHK$_{s}$, ][]{capak04} and our own K$_{s}$ imaging
data (Barro et al., in preparation). {\it Spitzer} IRAC data and MIPS
24~$\mu$m images were also used, jointly with GALEX observations in
the far ultraviolet (150~nm; FUV) and near ultraviolet (230~nm; NUV)
bands. In addition, we also used the {\em bviz} HST ACS imaging
covering the whole field.

For the EGS, we used the multi-wavelength dataset published by the
All-wavelength Extended Groth Strip International Survey (AEGIS, see
\citealt{davis07} for a detailed description). These data consist in
CFHT {\em ugriz} imaging, CFHT {\em BRI} \citep{coil04} observations,
{\em vi} HST-ACS data, {\it Spitzer} IRAC and MIPS images, and GALEX
FUV and NUV observations.

There is also a wealth of publicly available spectroscopic redshifts
in both fields. For the EGS, the DEEP2 Galaxy Redshift Survey
\citep{faber03} obtained over 15,000 redshifts in the whole EGS. In
the GOODS-N field, spectroscopy is available for $\sim$1,500 sources
\citep{Wir04,Cow04,Red06}.

\section{Sample selection}
\subsection{Color-magnitude diagram}
\label{col-mag}
\begin{figure}
\includegraphics[width=6cm, angle=-90]{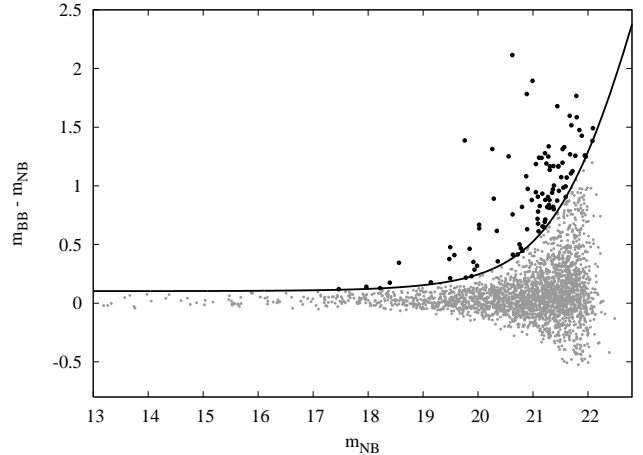}
\caption{\label{cm-diag} Color-magnitude diagram for the Groth3 field. 
Emission-line candidates are represented with black circles, and the
rest of the objects are represented with grey circles. The
2.5-$\sigma$ selection curve is also plotted. Fluxes are measured
within a four pixel diameter aperture.}
\end{figure}

Emission-line objects were selected by their excess flux when
comparing the narrow band and the broad band images. The candidates
were selected as those showing a clear flux excess. The criterion used
was:

\begin{equation}
(m_{BB} - m_{NB}) > \mu(m_{BB} - m_{NB}) + n_{\sigma} \; \sigma(m_{BB} - m_{NB})
\end{equation}

\noindent where $m_{BB}$ is the apparent magnitude in the broad-band filter, $m_{NB}$
is the apparent magnitude in the narrow-band filter, $\mu$ is an
offset parameter, i.e. the average deviation from the zero color,
$\sigma$ is the standard deviation of the color distribution, and
$n_{\sigma}$ is the level of significance. The offset parameter and
standard deviation can be expressed as a function of the narrow-band
magnitude, and they can be calculated directly from the distribution
of objects. Thus, we have a certain curve, dependent on the
narrow-band magnitude, above which objects are selected as
emission-line candidates. In Figure~\ref{cm-diag}, we show the
color-magnitude diagram with the selection curve for one of our
fields.

Fluxes in each band were measured within fixed circular apertures of
different sizes. Thus, our measurements have the same spatial origin
avoiding the mix of light from different regions in extended
galaxies. The selection of the candidates was carried out using
different apertures, typically ranging from the PSF FWHM to five times
this quantity, for a total of 9-10 apertures. 

The main goal in the selection process is to efficiently select
emission-line objects avoiding (redshift and non-emission-line)
interlopers. Taking advantage of the large spectroscopic surveys in
both the EGS and GOODS-N fields, we studied the the level of
significance and range of apertures that yielded better results.

\begin{figure}
\includegraphics[width=7.5cm,angle=-90]{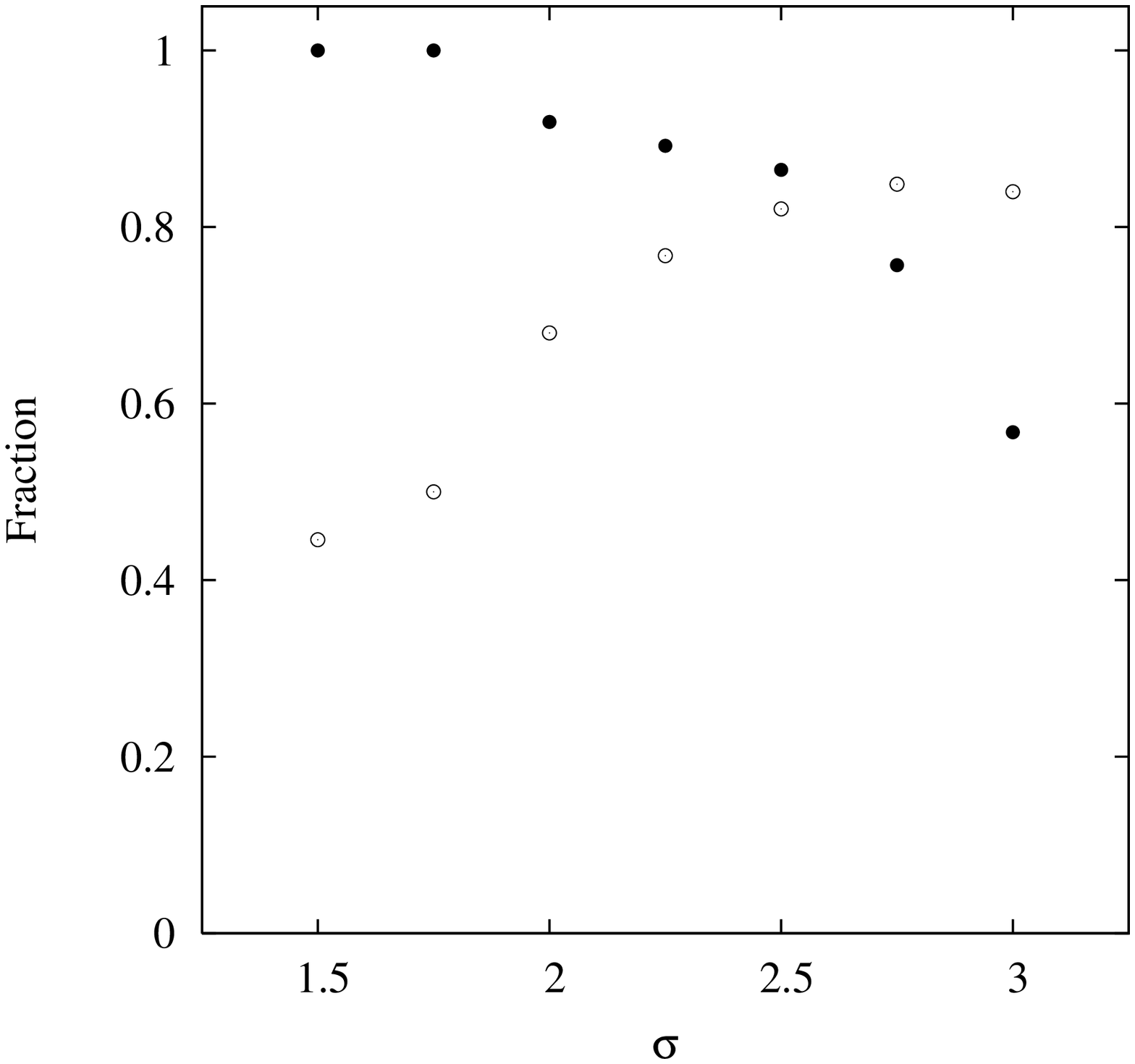}
\includegraphics[width=7.5cm,angle=-90]{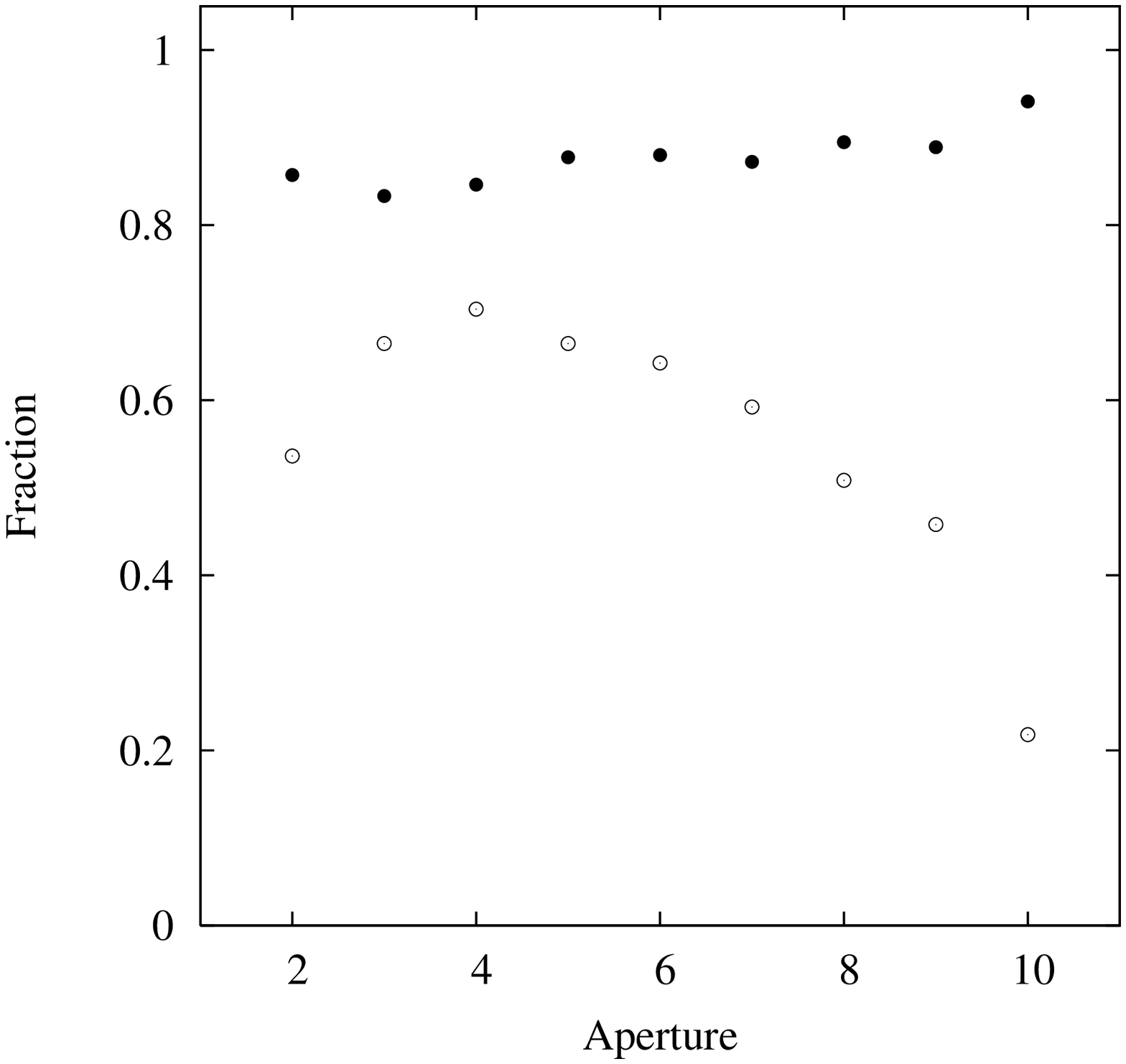}
\caption{\label{check_sigma} Left: Analysis of different significance
levels $n_{\sigma}$. Open circles represent the fraction of
emission-line objects in the total selected sample at each
$n_{\sigma}$. Filled circles represent the fraction of emission-line
objects selected at each $n_{\sigma}$ level, considering the total as
the number of objects selected at the lowest significance level,
i.e. $n_{\sigma}$=1.5. Right: Selection results in different
apertures. Open circles: fraction of total number of objects selected
at each aperture over the total number obtained with all the
apertures. Filled circles: fraction of confirmed emission-line
objects.}
\end{figure}

In order to study the best level of significance, we created several
selection curves with values of $n_{\sigma}$ ranging from 1.5 to 3.0
in steps of 0.25. Each selection curve defines a sample of emitting
candidates. We then obtained spectroscopic redshifts for each sample
by cross-matching our selected samples with spectroscopic
catalogs. The search radius was set to 1$\arcsec$. The objects with
spectroscopic redshift can be divided into those selected by an
emission-line (i.e. selected by H$\alpha$, [O{\sc
iii}]$\lambda\lambda$5007,4959 or [O{\sc ii}]$\lambda$3727 line flux),
and those not selected by any of these emission-lines. The fraction of
the former objects over the total tell us how accurately we are
selecting genuine emission-line galaxies. The final goal is to select
the maximum number of objects without losing accuracy. Using the
lowest significance level, we obtain a sample of candidates with the
largest number of objects, but many of them could be redshift
interlopers. Assuming that every object in the redshift range is an
emission-line object, we can measure the fraction of objects recovered
over the total in the lowest significance level. The best level of
significance will be a compromise between accuracy and number of
selected objects. In Figure~\ref{check_sigma}, we show the results for
each level of significance. We demonstrate here that a level of
significance $n_{\sigma}$=2.5 is a good compromise between the number
of selected emission-line objects and the accuracy of the selection.

To include as many different line emitters as possible, it is
necessary to use several apertures. The smallest apertures are more
adequate for the detection of small, low luminosity emission-line
objects, since the corresponding fluxes are less affected by the sky
noise. This is also the case for bright objects with high nuclear star
formation. On the other hand, large, low surface brightness objects
with extended star formation are better selected with the larger
apertures. Large apertures measurements are very noisy for small
objects, so apertures significantly larger than the object were not
considered. Figure~\ref{check_sigma} shows the fraction of objects
selected in each aperture over the total number obtained by taking
into account all the apertures. In addition, we represent the accuracy
at each aperture. If we select emission-line candidates in a
1.8$\arcsec$ (4 pixels) diameter aperture, we recover $\sim$70\% of
the objects in the final sample using all apertures. Thus, we are
losing $\sim$30\% of the objects if we only use one aperture, even if
it is the one that selects the highest number of objects. The accuracy
in each aperture remains constant, even for the larger apertures where
the sky noise could severely affect small objects fluxes. The reason
is that we reject those objects selected in apertures much larger than
its size.

\subsection{Star-galaxy segregation}

After selecting the candidates to be an emission-line at z$=$0.84, we
must determine if the source is a star or a galaxy. The discrimination
between stars and galaxies was carried out using eight different
criteria. The main criterium was the {\tt STELLARITY} parameter given
by Sextractor in each optical and NIR band were the object was
detected. Every object presenting an average value of the {\tt
STELLARITY} parameter higher than 0.95 was classified as a star.

In addition we used the following color criteria based on IRAC and NIR
magnitudes \citep{Eisen04,RR05}: a) $[3.6] - [8.0] > -2$ and
$[3.6]-[8.0] < -1$ and $[8.0] < 20.$, or $[3.6]-[4.5] > -1$ and $[3.6]
- [4.5] < -0.5$ and $[4.5] < 19.5$; b) $[5.8] - [8.0] > -1$, $[5.8] -
[4.5] < -0.2$ and $[8.0] < 20.$; c) $I - [8.0] < -1$ or $I - [3.6] <
1$ and $[3.6] < 18.$ or $I - [8.0] < -1$ and $[3.6] - [8.0] < - 1$; d)
$B - I > 2 × (I - [3.6]) + 0.070;$ e) $J - K + 0.956 < 0.5$; and f)
$[3.6]_{3\arcsec} - 0.460 - [3.6]_{auto} > -0.25$ and $[3.6] < 15.$
and $[3.6]_{3\arcsec} -0.460-[3.6]_{auto} < 0.2$, or $[3.6]_{3\arcsec}
- [3.6]_{auto} < - 0.25$, where [band]$_{3\arcsec}$ is the magnitude
in a 3$\arcsec$ diameter aperture, and [band]$_{auto}$ is the mag auto
magnitude given by sextractor (an estimation of the integrated
magnitude). The BzK criterion $(z-K)_{AB} <0.3\times (B-z)_{AB}-0.5$
\citep{daddi04} was also used.

Only 4 objects where classified as stars in the total sample of 243
candidates. Half of them were selected in the GOODS-N field and the
other half in the field Groth2. This represents 1.6\% of the whole
sample, a clearly negligible fraction.

\subsection{Photometric redshifts}

Once the stellar objects have been removed from the sample, we tried
to get rid of the objects outside the redshift range we are
studying. We showed in Section~\ref{col-mag} that we can have two
types of redshift interlopers in our sample: a) those selected by
other emission-lines; and b) those selected due to noise or strange
spectral features. Spectroscopic redshifts with enough quality were
available for 98 out of 239 objects (241 if we include the
stars). This means that there are 141 objects (59\% of the entire
sample) without spectroscopic data. Estimating photometric redshifts
(despite their relatively high uncertainties compared to spectroscopic
values) for these objects is important to get a highly complete and
reliable sample of galaxies at z$\sim$0.84.

\begin{figure}
\includegraphics[width=7.5cm]{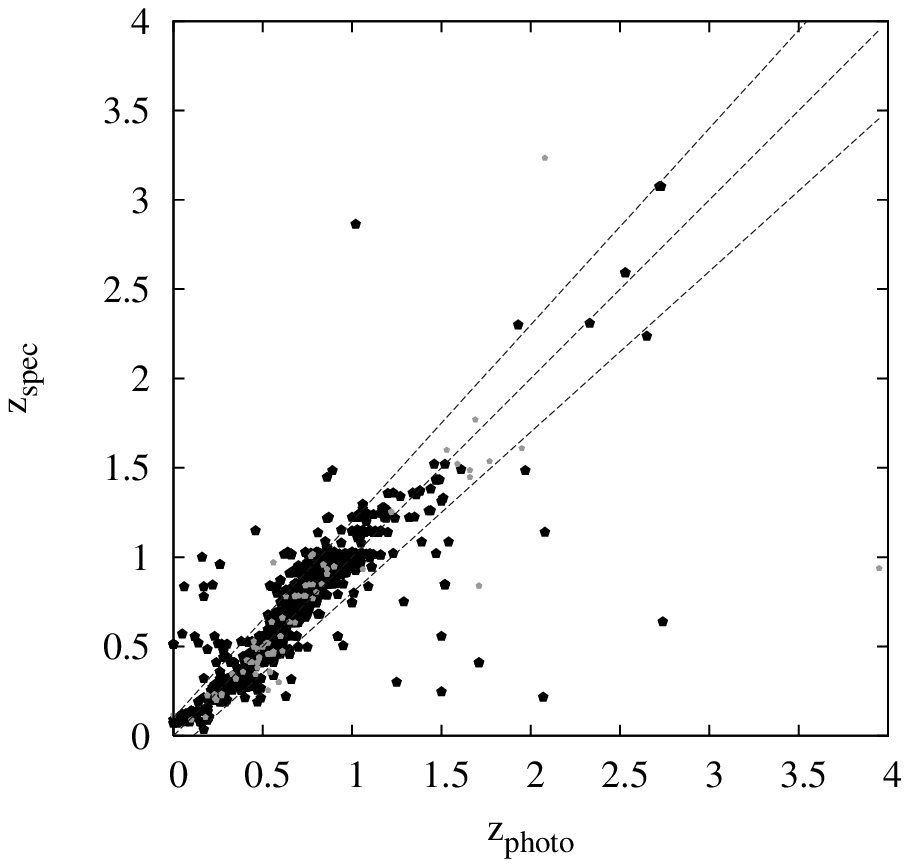}
\includegraphics[width=7.5cm]{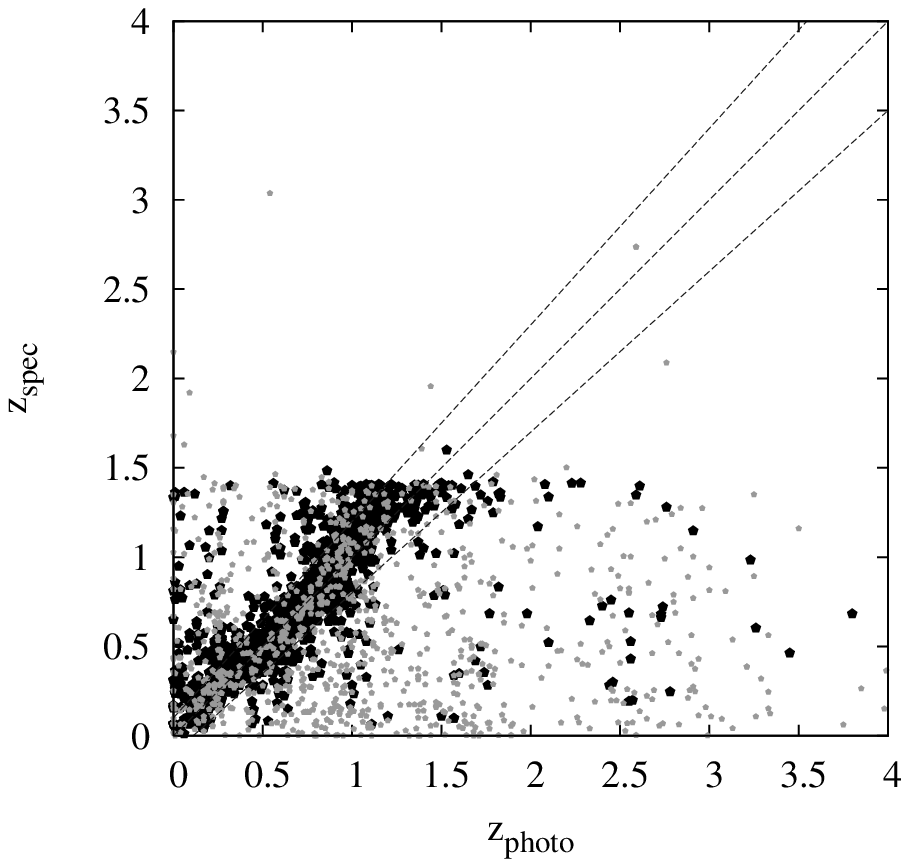}
\caption{\label{photo-z} Comparison between spectroscopic and photometric 
redshift for sources selected in the $J$-band in the GOODS-N (left
panel) and Groth fields (right panel). Black points are objects whose
spectroscopic redshift quality is very good. Grey points have
spectroscopic redshifts with low reliability flags. Note that,
although it seems to be a lot of dispersion in the EGS field, 86\% of
the objects with reliable spectroscopic redshift fall within
$\sigma_{z}/(1+z)<$0.1 (dark lines in the figure), and 95\% fall
within $\sigma_{z}/(1+z)<$0.2.}
\end{figure}

We obtained photometric redshifts for our sources using the same
method presented in \cite{PG05} and appendix B of
\cite{PG07}. First, we measured consistent (aperture matched) photometry 
in each band where the object was detected. Then, a set of templates
(built with stellar population and dust emission models) was
redshifted (in steps of $\Delta z=0.1$) and convolved with the
observed filters. A $\chi^{2}$ minimization algorithm was used to
estimate the most probable photometric redshift for each object. A
preliminary step determining the 1.6$\mu$m bump feature helped to
constrain the final solution. An additional constraint was imposed to
the template that best fitted the data points: it had to be younger
than the age of the universe at the given photometric redshift. The
photometric redshift probability distribution was built with the best
$\chi^{2}$ values for each redshift. This probability distribution was
very useful because some objects had two or even more peaks, making
them compatible with different redshifts.

\begin{figure}
\includegraphics[width=8.5cm]{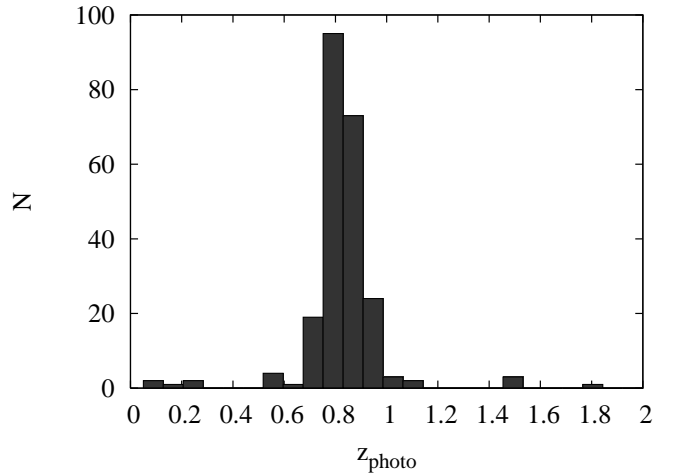}
\caption{\label{hist_photz08} Photo-redshift histogram for all 
galaxies in our three fields with reliable spectroscopic redshift
within our redshift range. Some of these objects have several peaks in
the probability distribution that move them to the central Gaussian
distribution.  }
\end{figure}

We estimated photometric redshifts for all the objects with
spectroscopic redshifts detected in the $J$-band
images. Figure~\ref{photo-z} shows the comparison between
spectroscopic and photometric redshifts for the GOODS-N and Groth
fields. In the first panel, we show the comparison for the 1430
$J$-band sources with available spectroscopy in GOODS-N. Although
there are some sources that lie quite far from the one-to-one
relation, most of them have a photometric redshift in good agreement
with the spectroscopic value. There is no evidence of a significant
systematic error, given that the average difference $\delta
z=z_{spec}-z_{photo}$ is 0.011, 90\% of
the objects with reliable spectroscopic redshift fall within
$\sigma_{z}/(1+z)<$0.1, and 97\% fall
within $\sigma_{z}/(1+z)<$0.2.

The results for the EGS fields are shown in the second panel, with a
total of 3810 sources. In this case, the quality of the
photo-redshifts is very similar to that achieved in GOODS-N. The
average difference between redshifts is $\delta z$=-0.011, 86\% of
the objects with reliable spectroscopic redshift fall within
$\sigma_{z}/(1+z)<$0.1, and 95\% fall
within $\sigma_{z}/(1+z)<$0.2.

Due to the typical photometric redshift uncertainties, the objects
with spectroscopic redshift in the redshift range of interest (the one
corresponding to the H$\alpha$ emission for our NB filter) are spread
over a much wider photo-redshift range. Figure~\ref{hist_photz08}
shows the histogram of photometric redshifts for the galaxies with
spectroscopic redshifts within our filter's range. Most of the sources
are close to the expected spectroscopic value of z$=$0.84 with some
outliers. However, some of these outliers presents peaks in the
probability distribution that shift them close to the spectroscopic
redshift. The mean (median) photo-redshift value of the distribution
plotted in Figure~\ref{hist_photz08} is z$_\mathrm{photo}$$=$0.822
(0.820), whereas z$_\mathrm{spec}$$=$0.839 (0.838). The difference is
$\delta z$$=$0.017 (0.018), and the standard deviation for the
photometric redshifts is 0.16.

Objects with a measured spectroscopic redshift outside the range
covered by the filter were removed from the sample. Sources with no
spectroscopic redshift and photometric redshift
z$_\mathrm{photo}$$<$0.5 or z$_\mathrm{photo}$$>$1.1 were also
discarded. Note that we checked the photo-z probability distribution
for each of these sources and, in the cases where there was a peak at
0.5$<$z$_\mathrm{photo}$$<$1.1, it was introduced again in the
sample. Finally, eight galaxies without spectroscopic nor photometric
redshift were kept in the final sample.

After the removal of stars and redshift interlopers, the final sample
of H$\alpha$ emitters at z$\sim$0.84 has 165 objects: 51 in Groth3, 56
in Groth2 and 58 in GOODS-N. Of these, 165 galaxies, 79 (58\%) are
confirmed spectroscopically. Table~\ref{tab_objects} lists the objects
in the final sample. The [O{\sc ii}]$\lambda$3727 and [O{\sc
iii}]$\lambda\lambda$4959,5007 emitters will be analyzed elsewhere.

\subsection{Survey detection limits and completeness}
\label{incomplete}
The narrow-band technique to select emission-line galaxies at
different redshifts has been extensively used over the last
years. However, most of the times, the line flux detection limit is
not consistently determined. The problem is that two different images
are used and that each line flux could come from different
combinations of narrow and broad band fluxes (i.e., galaxies could
cover a wide range in equivalent widths).

In this work, we decided to tackle this problem performing simulations
of the selection and measurement processes, in order to determine: a)
the completeness of our selection; b) corrections for incompleteness
in the luminosity function; and c) systematic errors that could lead
to erroneous line flux measurements.

The method consisted in introducing a well known sample of fake
galaxies in the science images and, working exactly in the same way as
we do with the real images, check whether or not we recover the
original properties of the fake sample.

\begin{figure}
\includegraphics[width=8.5cm]{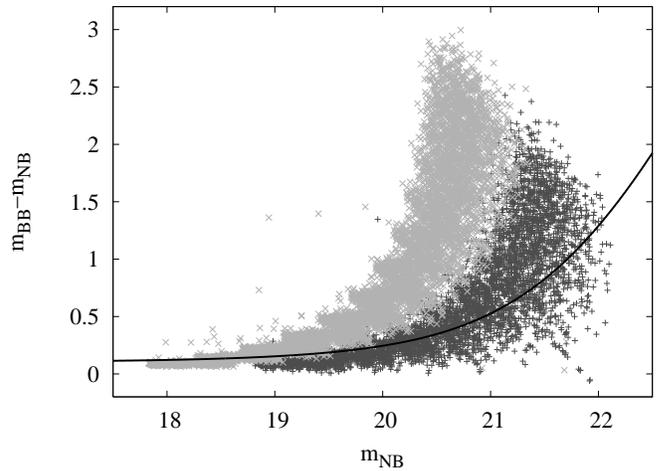}
\caption{\label{color_sim} Color- magnitude selection diagram of simulated 
objects in the Groth3 field. The dark-grey points have
f$_l$$=$10$^{-16}$~erg~s$^{-1}$~cm$^{-2}$ and the light-grey
ones have
f$_l$$=$1.5$\times$10$^{-16}$~erg~s$^{-1}$~cm$^{-2}$. All
magnitudes were measured in circular apertures for 4 pixels
diameter. The solid line is the selection curve for this field and
aperture size, with only objects above it considered as candidates
(for this aperture).}
\end{figure}

The analysis of the HST morphology of our sample (Villar et al, in
preparation) shows that most of our galaxies are disky, and that a
significant fraction of the global star formation (typically less than
50\%, with a mean value of 30\%) is distributed in several star
forming regions (being the mean number 5) covering the whole
galaxy. We used this information to model disks with star formation
distributed in five star forming regions, randomly distributed, within
the galaxy's half light radius. This produces models of galaxies with
highly concentrated as well as more extended and diffuse star
formation. We used a exponential law for the disks, limiting the
models to three different half light radius and three different
inclinations. The star forming regions were modeled with Gaussian
profiles and a half light radius of 600pc, which is the average radius
we found in the morphological study. In Table~\ref{tab_prop}, we give
the range of parameters covered by the models. The fake galaxy images
were constructed using {\tt GALFIT} \citep{Peng02}, convolving the
model with the field's PSF. No additional noise was added to the
models because the main source of noise for faint objects in our
images was the sky background level. For each combination of
parameters, we inserted 200 fake galaxies in the science image. We
then carried out the detection of candidates in the standard way.

\begin{deluxetable}{lr}
\tablecaption{Simulations: range of parameters}
\tablehead{
\colhead{Physical property} &\colhead{Range}
}
\startdata
Log Line flux(erg s$^{-1}$ cm$^{-2}$)& -15.1 -- -16.5\\
H$\alpha$ equivalent width (\AA)& 10 -- 2500\\
Effective radius (kpc) & 2.5, 5.0, 7.5\\
Inclination ($^\circ$)&0, 45, 70\\

\enddata
\label{tab_prop}
\end{deluxetable}

For a certain line flux, we have different narrow-band and broad-band
fluxes. The magnitudes for each band are given by:
\begin{eqnarray}
m_{NB}=C-2.5 \log (f_c+f_l/\Delta_{NB})\\
m_{BB}=C-2.5 \log (f_c+f_l/\Delta_{BB}) 
\end{eqnarray}

\noindent where $C$ is the zero-point, $f_c$ is the continuum flux 
per wavelength unit, $f_l$ is the line flux and $\Delta_{NB}$ and
$\Delta_{BB}$ are the narrow- and broad-band filter effective
widths. Objects with the same line flux present different narrow-band
magnitudes with a maximum value given by $m_{NB}=C-2.5 \log
(f_l/\Delta_{NB})$.  In Figure~\ref{color_sim}, the different
locations in the color-magnitude diagram are shown for objects with
the same flux level simulated in the Groth3 field. The fluxes were
measured in apertures of 4 pixels in diameter and the corresponding
selection curve is also shown. We can see that the color, i.e., the
equivalent width, increases with narrow-band magnitude. The reader
should note that low equivalent width objects are not selected by our
method. However, these sources also present bright $J$-magnitudes and
they are not relatively very numerous, so the completeness will not be
seriously affected. The fluxes for the faintest sources are recovered
with less accuracy and the dispersion becomes larger, preventing the
selection of the whole fraction of objects.

\begin{figure*}
\epsscale{.10}
\includegraphics[width=8.0cm]{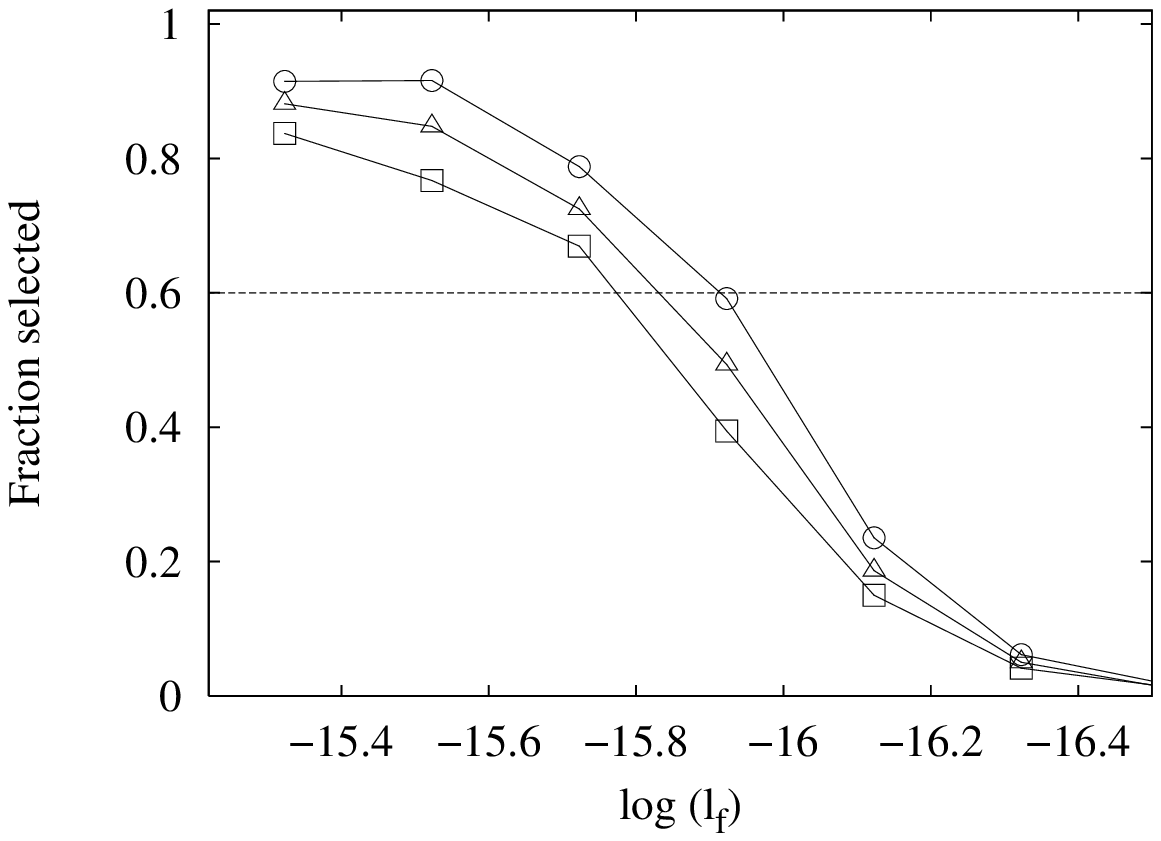}
\includegraphics[width=9.0cm]{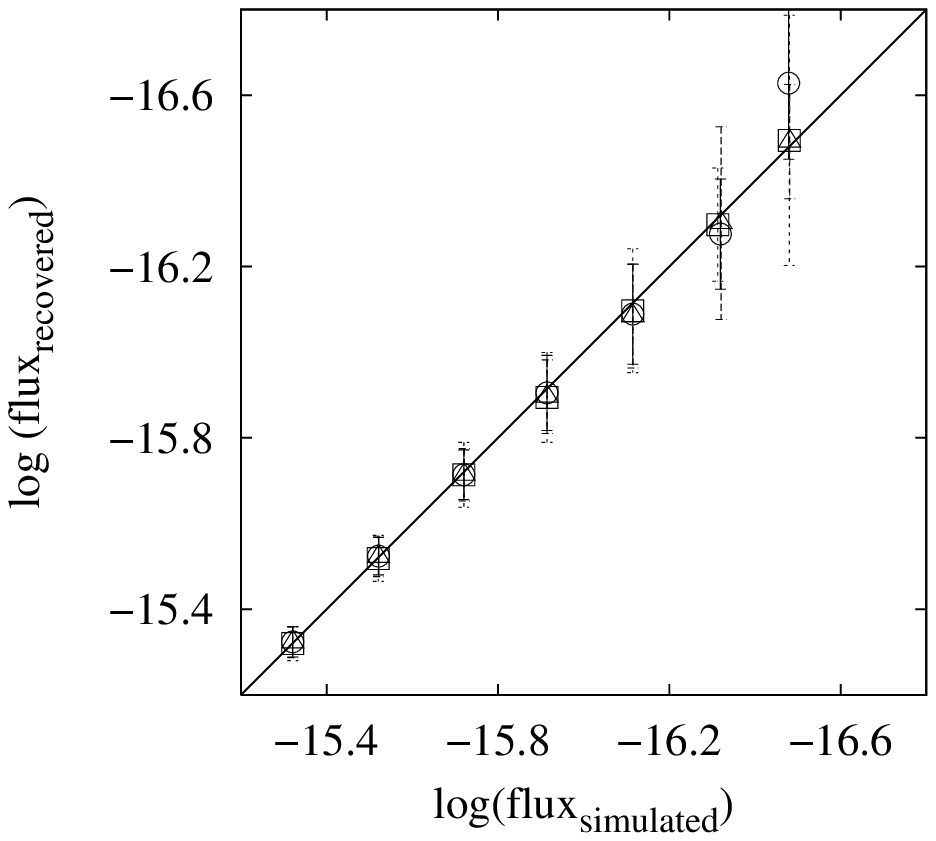}
\includegraphics[width=8.0cm]{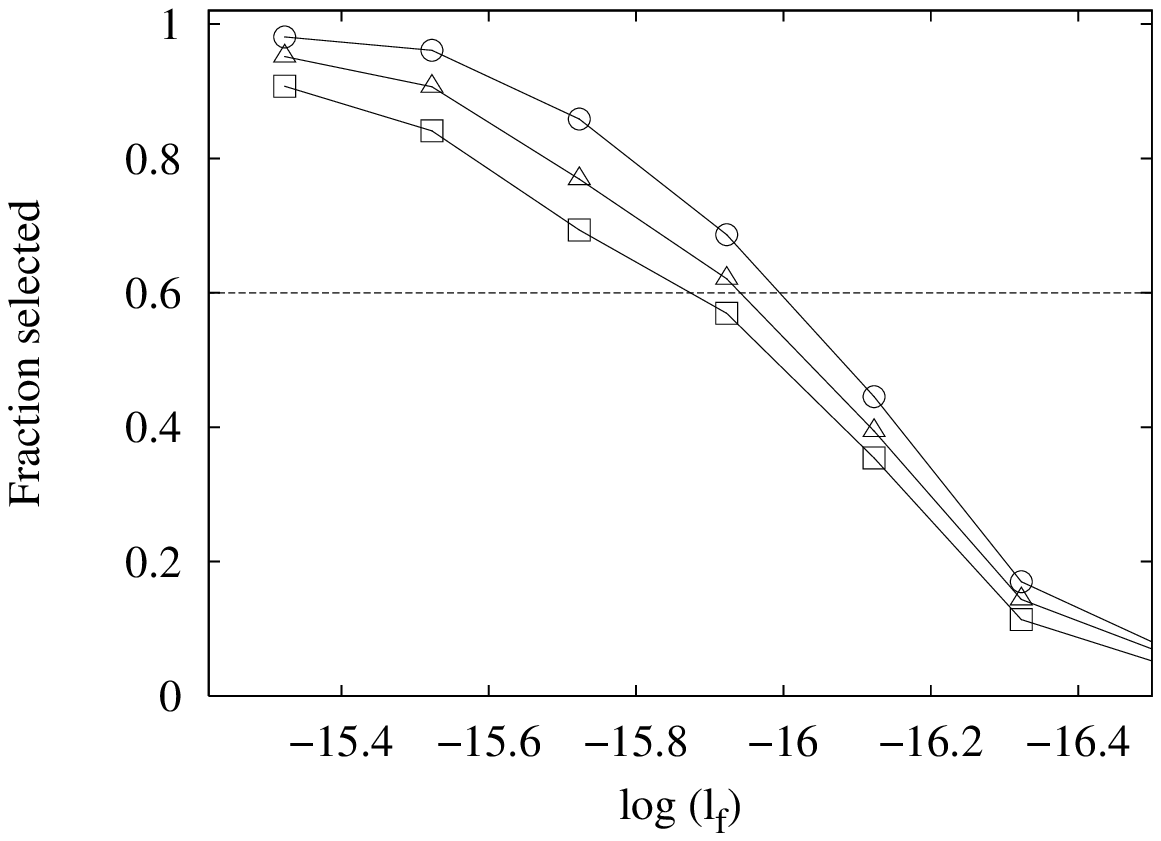}
\includegraphics[width=9.0cm]{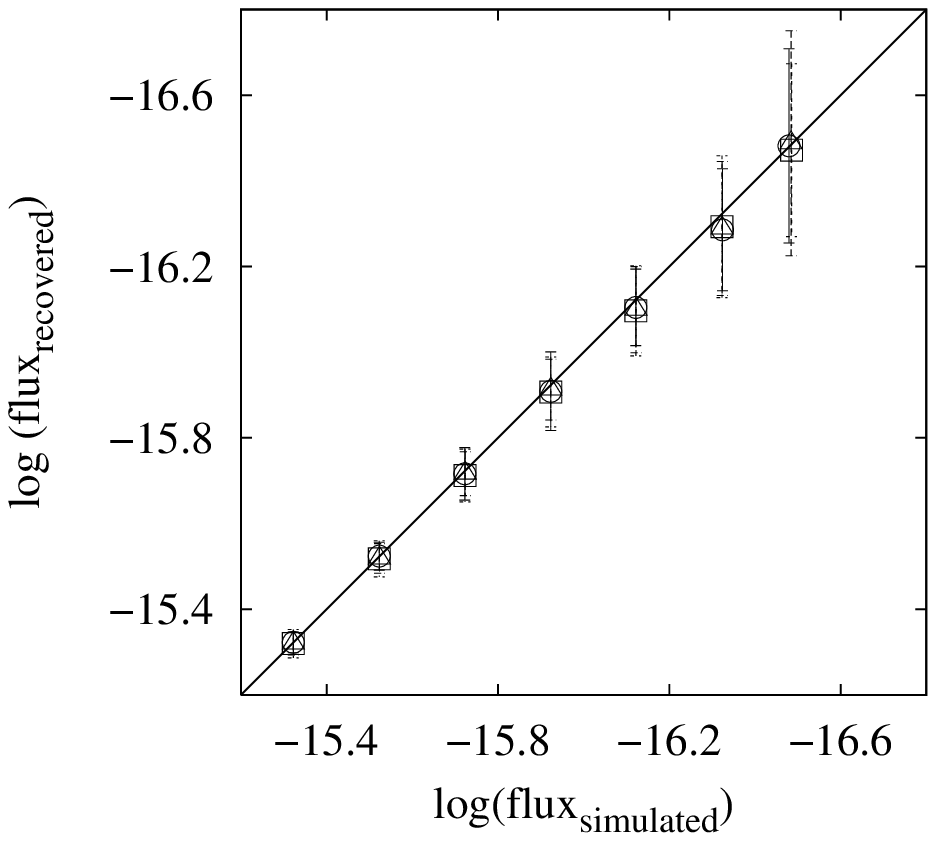}
\includegraphics[width=8.0cm]{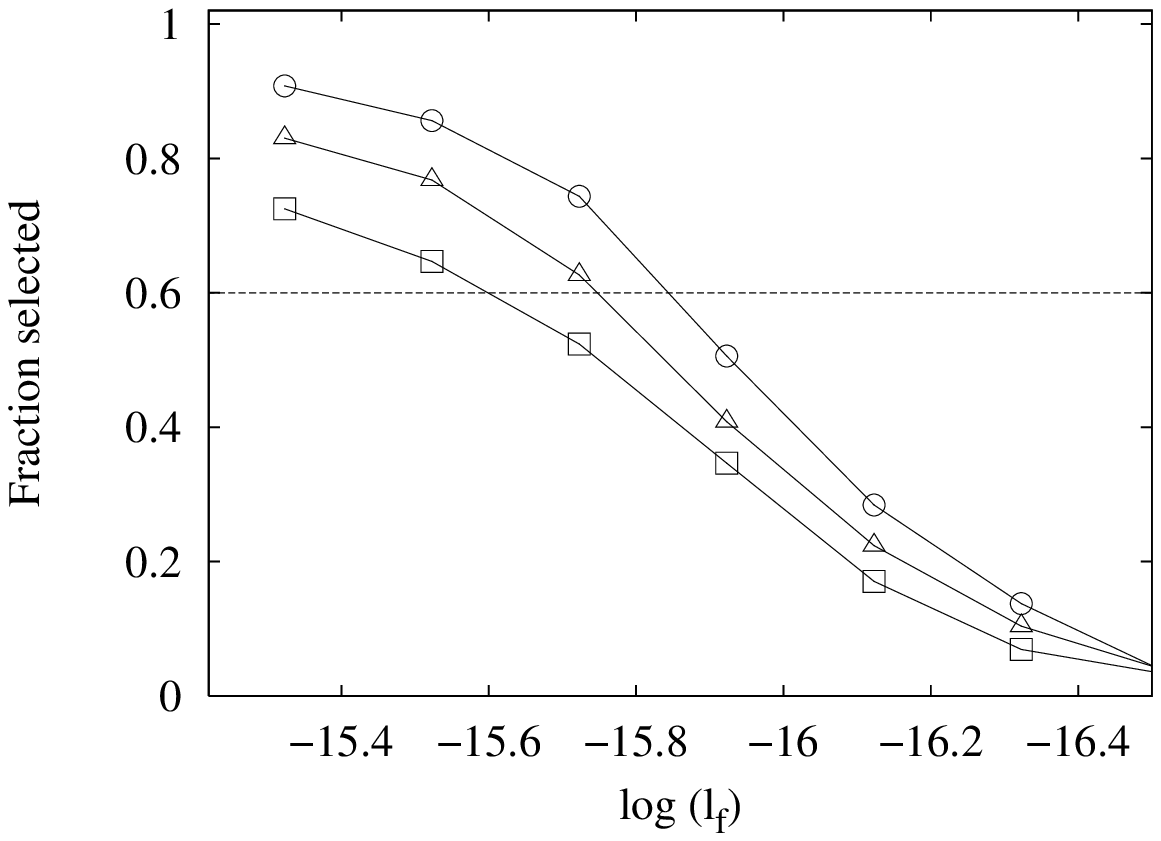}
\includegraphics[width=9.0cm]{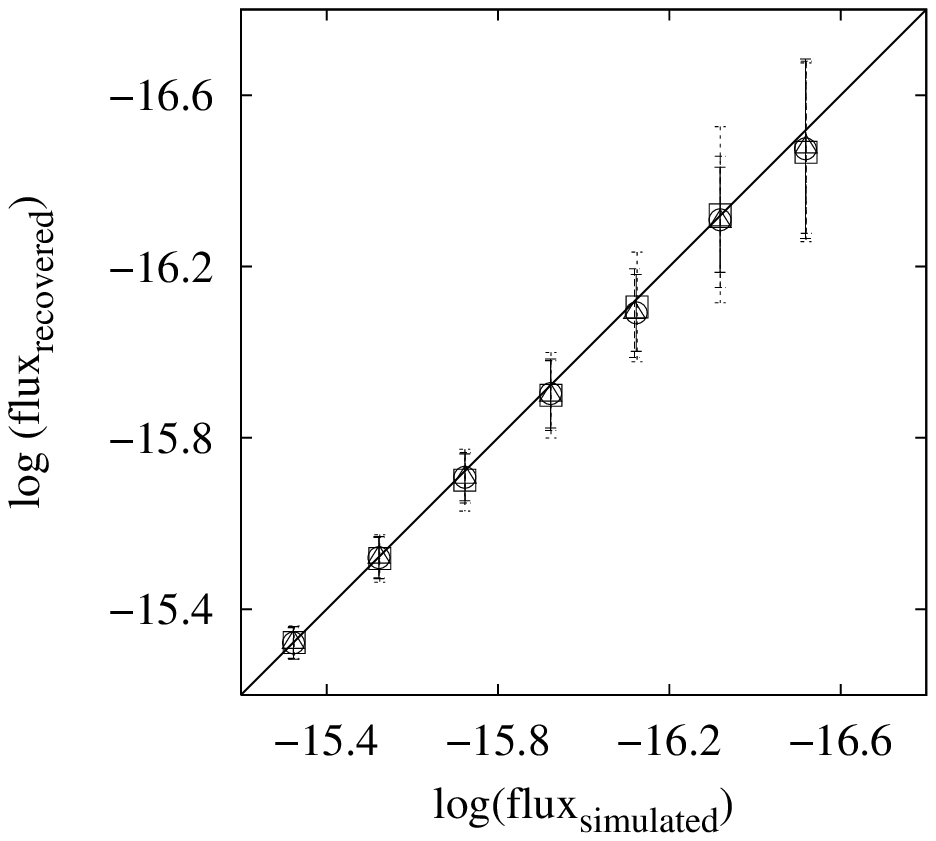}
\caption{\label{simu1} Completeness and line flux accuracy for each field surveyed. 
From top to bottom: Groth2, Groth3, GOODS-N. Left: Completeness versus
line flux. Different symbols represent different half-light radius of
the simulated sources: circles for 2.5 kpc, triangles for 5 kpc and
squares for 7.5 kpc (assuming z=0.84). Right: Recovered flux versus
input line flux. Symbols are the same as in the left panels. }
\end{figure*}

In addition, Figure~\ref{color_sim} shows that there is an upper limit
to the color that decreases with narrow-band magnitude. The
explanation is that high equivalent widths imply very faint fluxes in
the broad-band image. Consequently, these objects would not be
selected by our technique, since we need a simultaneous two band
detection to assure the existence of the source. It is possible,
however, to use the narrow-band image to detect all objects and then
measure at the same position in the broad-band image. We did not apply
this method because the alignment of the images was not good enough
throughout the whole image, producing wrong centered apertures in the
broad-band images in some regions, which lead to incorrect results. A
refinement of the alignment process could not be done without
substantial transformation of the images, which could alter the final
results. We conclude that we may have missed faint galaxies with large
equivalent width values undetected in the $J$-band. In order to
analyze this systematic detection effect, we checked the images
looking for objects only detected in the narrow-band images. We could
not find any reliable candidate. Moreover, the maximum observed
equivalent width measured in our sample is 1077\AA, with the rest of
the sample below 600\AA. This is in good agreement with the results
found in \cite{Gallego95}, \cite{Tresse02}, and \cite{P05}. A deep
narrow band survey looking for Ly$\alpha$ emitters at redshift z=8.8
\citep{wil05} did not find any population of high EW(H$\alpha$)
emitters. However, this survey was carried out over a small
area. Our survey covers a much wider area and confirm these previous
results.

Figure~\ref{simu1} (left) shows the fraction of selected objects for
the different surveyed fields. Each panel correspond to a different
field and shows the completeness for different half light radius. The
completeness curve shows a smoother decline than that we can found in
magnitude completeness studies. The reason is that in the narrow-band
technique, two different magnitudes are involved and for each line
flux we span a wide range in broad- and narrow-band magnitudes (see
Figure~\ref{color_sim}). Another important issue is the effect of
increasing the half light radius. The completeness drops from
$\sim$80\% to $\sim$50\% when we move from r$_{eff}$=2.5 kpc to
r$_{eff}$=7.5 kpc in the GOODS-N field at $\log$(l$_{f}$)$\sim$-15.65,
and would move to lower fractions for higher half light
radii. However, this is not a major concern in our case, since 85\% of
galaxies present half light radius lower than 7.5 kpc, with all of
them except one below 10 kpc (Villar et al., in preparation).

The simulations allowed us to check the reliability of the measured
fluxes.  For the objects that satisfy the selection criteria, there is
a good agreement between the mean recovered value and the mean
simulated flux (see Figure~\ref{simu1}, right panel), even for the
faintest line fluxes. The comparison between individual objects in
each line flux bin give us a better estimation of the error than that
determined with photometric errors. Error bars in figure~\ref{simu1}
show the standard deviation in the recovered line flux, computed as
the standard deviation of the absolute difference between recovered
and simulated line fluxes. The errors clearly increase as we move to
fainter line fluxes, ranging from a 10\% relative error for the
brightest objects to a 60\% for the faintest ones, although they keep
below 30\% up to f(H$\alpha$+[NII])$=$5$\times$10$^{-17}$ erg s$^{-1}$
cm$^{-2}$.

\section{H$\alpha$ luminosities for z$\sim$0.84 objects}
\subsection{Line flux estimation}

Emission-line fluxes were computed using:
\begin{equation}
f_{\mathrm{l}} =  \Delta_{\mathrm{NB}} \left(f_{\mathrm{NB}}-f_{\mathrm{BB}}\right)\frac{1}{1-\epsilon}
\end{equation}

\noindent where $\ensuremath{f_{\mathrm{NB}}}$ and $\ensuremath{f_{\mathrm{BB}}}$ 
are total fluxes in the narrow- and broad-bands, $f_{\mathrm{l}}$ is
the line flux (including [NII]$\lambda\lambda$6548,6584),
$\Delta_{\mathrm{NB}}$ is the width of the narrow-band filter computed
following the procedure specified by \cite{P07}, and $\epsilon$ is the
ratio of the widths of the narrow- and broad-band filters.

To estimate the integrated emission-line flux of each galaxy, we used
the whole set of apertures. The flux grows with aperture diameter
until the end of the emission region or the sky are reached. Since
galaxies present a variety angular sizes, apertures of different sizes
must be used. For small objects, the maximum flux will be reached in a
small aperture, whereas for large objects it will be reached in larger
apertures. We visually checked the emission-line fluxes for each
aperture in each galaxy to select the more reliable integrated
emission-line flux.

\begin{figure}
%\epsscale{.10}
\includegraphics[width=9.cm]{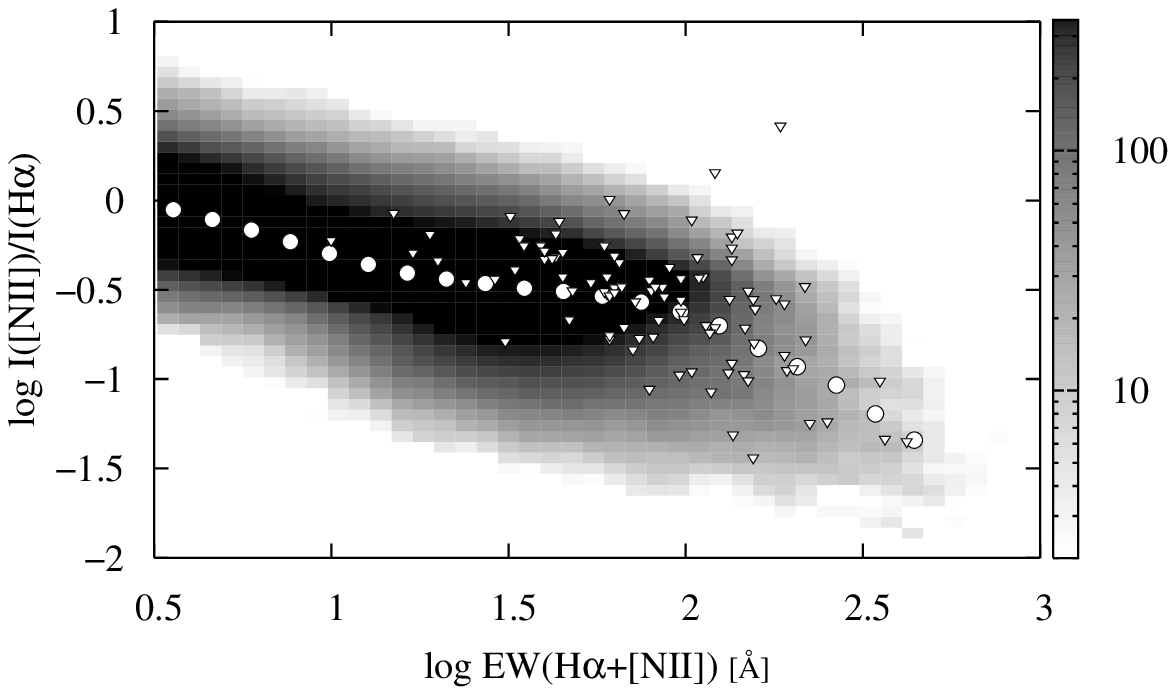}
\caption{\label{nitro_fig} Ratio $I[NII]\lambda6584/I(H\alpha)$ 
as a function of EW(H$\alpha + [NII]\lambda6584)$ for the Sloan Digital Sky
Survey (the number of galaxies is represented in gray scale) is represented in gray scale and the UCM survey (inverted
triangles). The mean values for the SDSS are represented as circles.}
\end{figure}

H$\alpha$ luminosities were computed from the line fluxes. The
underlying stellar absorption for H$\alpha$ has a negligible effect
when compared with errors from photometry, so no correction was added
\citep[see][]{nakamura04}. Nitrogen contamination to the narrow-band
flux was removed following the approach in \cite{P07}. In that work,
the shape of the narrow-band filter is considered when computing the
average [NII] contribution to the measured flux, assuming a certain
$I([NII]\lambda6584)/I(H\alpha)$ value. These authors assumed an
average ratio $I([NII]\lambda6584)/I(H\alpha)$=0.32, the mean value
obtained for the UCM Survey sample \citep{Gallego97} and the galaxies
in the Sloan Digital Sky Survey Release 4 \citep[SDSS
DR4][]{adel06}. In our case, we have used the SDSS DR4 to study the
dependence of $I([NII]\lambda6584)/I(H\alpha)$ with the equivalent
width of H$\alpha$ plus the [NII] contribution
[$EW(H\alpha+[NII])$]. Figure~\ref{nitro_fig} shows $\log
(I([NII]\lambda6584)/I(H\alpha))$ versus $\log
(EW(H\alpha+[NII]\lambda6584))$. There is a clear trend of decreasing
$I[NII]\lambda6584/I(H\alpha)$ as we move to higher equivalent widths,
which can be explained due to a metallicity decrease. The circles
represents the mean values obtained from the SDSS sample. The
dispersion remains at $\sim$0.4 dex for equivalent widths below $\log
(EW(H\alpha + [NII]\lambda6584))$=2. For higher equivalent widths, the
dispersion increases up to $\sim$1 dex. This relation was used to
estimate the [NII] contribution to the emission-line flux measured in
the narrow-band images for each of our sources, obtaining a mean
(median) value of $I[NII]\lambda6584/I(H\alpha)$=0.26 (0.27), ranging
from 0.04 to 0.4.

The $I[NII]\lambda6584/I(H\alpha)$ may evolve with redshift as galaxy populations could be very different in the past than local ones. However, changing this ratio by a factor of 2 implies $\sim$20\%-30\% variation in the H$\alpha$ line fluxes, which is of the order of the errors.

\subsection{Reddening correction}
\label{extinction}

Following \cite{buat05}, we used the ratio ($F_{dust}/F_{FUV}$) to
compute the extinction in the ultraviolet. Dust emission is given by
L(8-1000$\mu$m) and can be estimated for the objects detected by MIPS,
which traces the rest-frame continuum at 13$\mu$m in our redshift
regime. To carry out this estimation, we first subtracted the stellar
emission predicted by the stellar population templates (obtained in
the photo-z determination) from the fluxes at rest-frame wavelengths
redder than $\sim$4$\mu$m to obtain the pure emission of the
dust. Then, we fitted this emission with \cite{cha01} dust emission
models. The model that best matched the observed dust emission colors
[observed F(24)/F(8)] was selected, and we computed the TIR luminosity
L(8-1000$\mu$m) from this model \citep[for more details, see][]{PG07}.
The stellar population template, convolved with the FUV filter
transmission curve, also give us the FUV rest-frame flux. With the
$F_{dust}/F_{FUV}$ ratio, we compute A(FUV) and then the extinction in
H$\alpha$ applying the Calzetti extinction law \citep{cal00}, assuming
that the attenuation of the stellar emission is 0.44 times the
attenuation of the nebular emission. This law was empirically obtained
from local starburst with star formation rates of up to a few tens
M$_{\odot}$ yr$^{-1}$, very similar to our galaxies.

\begin{figure}
\includegraphics[angle=-90,width=8.5cm]{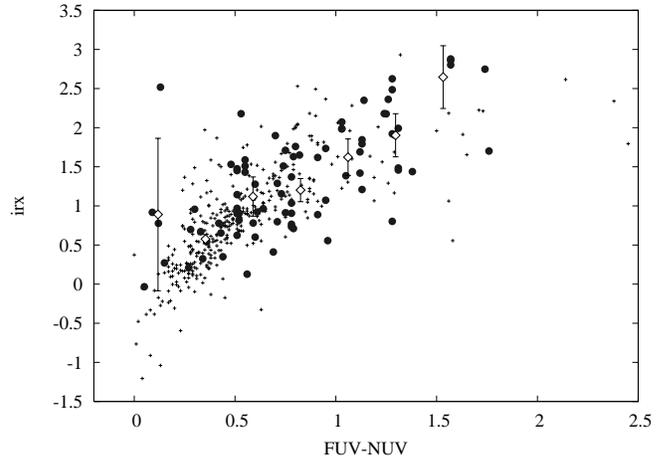}
\caption{\label{galex_fig} Dust flux to FUV flux ratio (irx) 
as a function of FUV - NUV color, i.e. the UV slope. The crosses are
the late-type galaxies in the GALEX Nearby Galaxy Atlas
\citep{Gil06}. Circles are objects in our sample that are detected at 24~$\mu$m
by MIPS. Rhombus show the mean values and dispersion for our sample. }
\end{figure}

However, 86 objects were not detected at 24$\mu$m, not allowing us to
obtain the dust flux. In this case, we approached the problem from the
ultraviolet side. The slope in the ultraviolet is another tracer of
the dust obscuration and it is correlated with $F_{dust}/F_{FUV}$
ratio, as \cite{Meu95} found for starburst galaxies. More recently,
\cite{Gil06} showed that the (FUV - NUV) color, which relates to the
UV slope \citep[see][]{kong04}, is also correlated with the
$F_{dust}/F_{FUV}$ ratio. The FUV - NUV color was computed convolving
the best stellar population template with the appropriate filter
transmission curves. Then, we estimated the $F_{dust}/F_{FUV}$ ratio
using the GALEX Ultraviolet Atlas of Nearby Galaxies. Each source in
our sample was assigned the mean GALEX atlas $F_{dust}/F_{FUV}$ at the
same (FUV -NUV). Figure~\ref{galex_fig} shows $F_{dust}/F_{FUV}$
vs. (FUV -NUV) for our 79 objects with MIPS detections. Late-type
galaxies in the GALEX atlas have also been represented. Our sample
follows, with higher dispersion, the general trend of nearby galaxies,
although they are, in general, redder than the local sample,
indicating that the extinctions are higher than those of the GALEX
atlas sample. This also indicates that there is little evolution of this relation with redshift, not having a significant effect on our results. Moreover, considering the objects with extinctions available by both methods, we obtain similar mean extinctions: A(H$\alpha$)=1.67~mag and A(H$\alpha$)=1.87~mag using the UV slope and the infrared excess respectively.

The mean extinction in our sample is A(H$\alpha$)=1.48~mag, a value
$\sim$0.5~mag higher than the mean values obtained for the SDSS
\citep{Brinchmann04} and UCM \citep{Gallego95} samples. This
implies an increase in the typical extinction of star-forming galaxies
with redshift of 0.5~mag from the local Universe to
z=0.84. 

 \cite{tresse2007} found that the dust obscuration at 1500\AA\
was A(FUV)=2 mag from z=0.4 to z=2, decreasing to $\sim$0.9--1~mag for
z$<$0.4. Our sample has a mean value of A(FUV)=2.15, in good agreement
with these authors. They argue that the decrease in extinction at low
redshift is due to the change of the dominant galaxy population. They
show that the emission from early-type galaxies starts to dominate in
the $B$-band below z$<$0.5, and they make the assumption that in the
FUV the early-type population will still dominate. Therefore, as the
dust content in early-type galaxies is much lower than in late-types,
the amount of extinction will decrease as we move to lower redshifts.

However, the difference we find when comparing the mean extinction of
our sample with that of local samples of star-forming galaxies cannot
be explained with that argument. \cite{Brinchmann04} showed that only
12\% of the SFR density comes from galaxies with D4000$>$1.8, and only
2\% from galaxies with D4000$>$2. Thus, only a very small fraction of
the star formation can be located in old systems with very poor dust
content. Moreover, \cite{Brinchmann04} pointed out that these systems
with high D4000 are probably spiral systems with significant
bulges. In addition, \cite{PG01} and
\cite{Vit96} did not find any elliptical galaxy
in the UCM sample and only 7\% of lenticular objects. So, taking into
account the previous discussion and the fact that our sample is
dominated by disks, thus sharing the morphology of the SDSS and UCM
samples, the higher extinction in our sample has to be caused by an
increase in the dust content in the galaxies that host the star
formation.

\section{The H$\alpha$ luminosity function at z=0.84}
\label{sec_lf}
\subsection{The observed H$\alpha$ luminosity function}
\label{ha_lf}
The H$\alpha$ luminosity function was calculated applying the
V/V$_{max}$ method \citep{Sch68}:

\begin{eqnarray}
\label{lumfunc}
\phi (\log L_i)=\frac{1}{\Delta \log L}\sum_{j} \frac{1}{V(z)_j}\\
%\nonumber \text{with} | \log L_j - \log L_i | < \frac{\Delta \log L}{2}
\end{eqnarray}

\noindent where $L_i$ is the central luminosity in the bin $i$ and $V(z)_j$
is the maximum volume in which object $j$ can be detected.

To properly compute the volumes defined by the narrow-band filter, we
followed the procedure described in section 5.3 of \cite{P07}. These
authors consider the volume in which an object would be detected in a
narrow-band survey based on its position in the color-magnitude
diagram, expanding the method to cope with several lines inside the
narrow-band filter. The effect of the Nitrogen lines become important
when the filter's transmittance falls and the H$\alpha$ line is
detected there. In that case, one of the Nitrogen lines would be in
the high transmittance region of the filter, increasing the total flux
and, hence, the detection probability. To include this contribution,
we considered the width of the filter affected by the Nitrogen lines
as in Equation (34) of \cite{P07}. If we consider and average volume
determined only by the narrow-band filter's FWHM, we would be
overestimating the surveyed volume by $\sim$20\% ($\sim$18\%) on
average (median), leading to a similar underestimation (i.e., a
systematic error) of the LF points. It is also important to take into
account the Nitrogen lines in the volume determination for each
individual object. If not considered, volumes are subestimated by
$\sim$40\% ($\sim$20\%) on average (median).

\begin{figure}
\includegraphics[width=8.5cm]{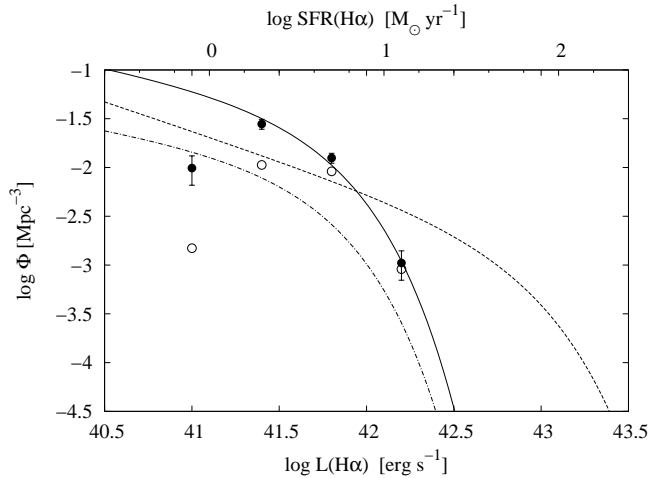}
\caption{\label{lf_nored} H$\alpha$ luminosity function not 
corrected for extinction (solid circles) with the best fit to a
Schechter function (thick line). Open circles represent the derived LF
before applying the completeness correction. for comparison
\cite{Tresse02} LF (dot-dashed line) and \cite{Hop00} LF (dashed line)
not corrected for extinction are also shown. }
\end{figure}

The H$\alpha$ luminosity function with no extinction correction is
shown in Figure~\ref{lf_nored}. The best fit to a \cite{Sch76}
function yields the following parameters:

\begin{eqnarray*}
\nonumber \phi^*=10^{-1.74\pm0.11} {\rm Mpc}^{-3}\\
\nonumber L^*=10^{41.69\pm0.07} {\rm erg}\ {\rm s}^{-1} \\
\end{eqnarray*}

We fixed $\alpha$=-1.35 (based on \citealt{Tresse98} and
\citealt{Shi07}) as our LF did not reach faint enough luminosities
to accurately determine it.

Errors were obtained from simulations. We computed a large number ($\sim$1000) of LFs, randomly changing the line flux for each object within a gaussian distribution, with $\sigma$ determined by the object line flux error. The final errors in the LF are the standard deviations of the distributions obtained from the simulations. We apply this same method for the errors in the Schechter fit. We did simulations varying the LF within the errors distributions, obtaining distributions the Schechter parameters. The final errors in these parameters are the standard deviations of these distributions.

Figure~\ref{lf_nored} depicts the expected distribution of observed line
fluxes for the H$\alpha$ line (since the Nitrogen correction was
already applied). To correct for incompleteness, we computed the
fraction of galaxies detected and selected at a certain line flux
level (Section~\ref{incomplete}), what we call the completeness
fraction. Then, we assumed that this fraction was the probability for
a galaxy with these properties to be detected and selected in our
sample. Thus, for each selected galaxy, we would expect the inverse of
the completeness factor to be the real number density of
galaxies. This is equivalent to multiplying each source's detection
volume by its completeness factor. Thus, in the LF computation, we
multiplied the detection volume of each galaxy by the completeness
factor. The LF corrected for incompleteness is shown in
Figure~\ref{lf_nored}. The correction is more severe as we move
towards fainter luminosity bins. It is very strong in the faintest
bin, but still it is most probably underestimated.

The LFs (not corrected for extinction) published by \cite{Tresse02}
and \cite{Hop00} are also shown in Figure~\ref{lf_nored}
\citep[converted to the cosmology used in this work,
see][]{Hop04}. \cite{Tresse02} observed a sample of galaxies at
z$\sim$0.7 selected from the Canada-France redshift survey with
EW([O{\sc ii}]$\lambda$3727])$\geq$12\AA.  Our LF is very similar to
Tresse's, although ours extends to higher luminosities. Also, our LF
presents a higher density at the faint end. This could be due to the
fact that \cite{Tresse02} applied a global completeness correction
independent of the line flux. This was the best they could do, since
they could not select the objects directly by their H$\alpha$
equivalent width or flux, but by [O{\sc ii}]$\lambda$3727 equivalent
width. A global completeness correction make the whole LF move to
higher densities whereas a flux dependent completeness correction
change the shape of the LF, raising the faint region. There is also
another caveat in their selection: not all the H$\alpha$ emitters show
[O{\sc ii}]$\lambda$3727\AA\ emission. \cite{Yan06} showed that in the
SDSS survey, 20\% of all emission-line galaxies with an H$\alpha$
detection have no [O{\sc ii}]$\lambda$3727\AA\ emission. In this
sense, if targets are selected in a spectroscopic survey using the
oxygen line equivalent width, a considerable fraction of an H$\alpha$
selected sample would not be detected. On the other hand, $\sim$30\% of
emission-line galaxies show oxygen emission with very low H$\alpha$
emission (22\%) or no emission at all (8\%).

\cite{Yan99} and \cite{Hop00} used the slitless spectroscopy technique 
to study emission-line galaxies at z$\sim$1.
\cite{Yan99} selected 33 emitters at 0.75$\leq$ z $\leq$ 1.9. Their 
data was not deep enough to constrain $\alpha$, so they assumed
$\alpha$=-1.35. \cite{Hop00} extended the study adding their deeper
data to that of \cite{Yan99}. The LF was similar to that of
\cite{Yan99}, although steeper. There is a huge discrepancy between 
our LF and theirs in the bright end of the LF, as they found many more
brighter objects. \cite{Tresse02} pointed out that, to some extent, it
could be an effect related to the Nitrogen correction. Indeed, the
slitless spectroscopy did not allow a proper deblending of the
H$\alpha$ line from the [NII]$\lambda$6548,6584 lines. However, we
have the same problem and we estimate a lower density of high
luminosity objects as well. Two explanations are possible: a
change in the shape of the LF at higher redshifts or field to field
variantions. The redshift range surveyed in
\cite{Yan99} and \cite{Hop00} is much larger than ours,
reaching higher redshifts (0.75$\leq$z$\leq$1.9). Star-forming
galaxies at z$\geq$1.4 could be very different from those at
z=0.8. For example, $\sim$30\% of the H$\alpha$ emitters at z$\sim$2
studied by \cite{Erb06} have $\log$(L$_{H_{\alpha}})>42.5$, whereas
our whole sample have lower luminosities. On the other hand, in the
volume surveyed by these authors there could be a high density region
due to cosmic variance. Probably, both effects are playing a role in
the comparison.

\subsection{The reddening corrected H$\alpha$ luminosity function}
\label{cor_lumfunc}

\begin{figure}
\includegraphics[width=8.5cm]{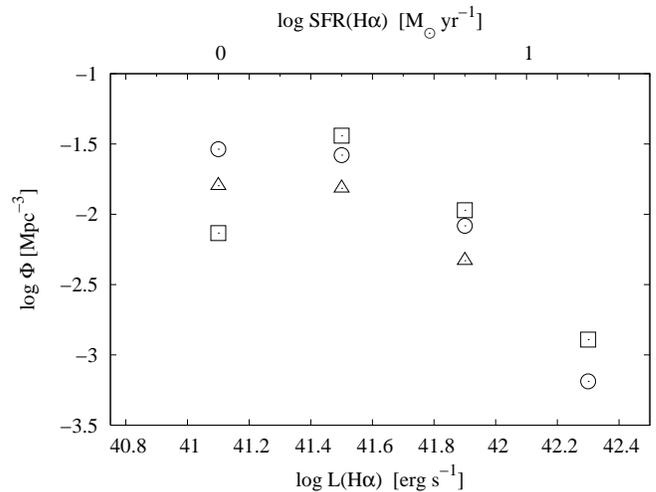}
\caption{\label{f2f} Derived LFs for the Groth2 (squares), 
Groth3 (triangles) and GOODS-N fields. The more populated points in
the central region shows a density ratio of $\sim$2.3 and $\sim$1.7
for Groth2 and GOODS-N fields over Groth3 field.}
\end{figure}

Two major effects are affecting our sample: extinction and field to field variance. The extinction correction was applied to each individual object and was explained in Section~\ref{extinction}. Field to field variance implies galaxy density changes depending on the observed field. Within our three surveyed fields, we notice significant field to field variations. Figure~\ref{f2f} show the different LFs computed for each field. The Groth2 and GOODS-N fields show an overdensity over the Groth3 field. If we limit the comparison to the bins log(L$_{H\alpha}$)=$\lbrace 41.5, 41.9 \rbrace$, which are less affected by low number statistics, the density of objects is $\sim$2.3 and $\sim$1.7 times higher in Groth2 and GOODS-N than in the Groth3 field, respectively. \cite{Taka07} reported a similar variation among the COSMOS and Subaru Deep (SDF) fields for their [O{\sc ii}]$\lambda$3727 emitters. In addittion, we notice that none of our fields could be representative of the mean density of star forming galaxies in the Universe at this redshift.

In order to correct for the field to field variance effect we use photometric redshift and spectroscopic redshifts to estimate: a) the mean density of our fields with respect to other fields and b) the relative density of galaxies within our redshift range over this mean density.

We used photometric redshifts (\citealt{PG07}, P\'erez-Gonz\'alez et
al., in prep.) for the EGS, GOODS-N, Chandra Deep Field South (CDFS)
and Lockman Hole (LH) to estimate the mean density in a redshift range
centered at our redshift, ranging from z$=$0.75 to z$=$0.9. We used a
redshift range wider than the narrow-band filter redshift range
because photometric redshifts do not work properly in such a small
range. We found that there was an overdensity of galaxies in both EGS
and GOODS-N fields. The overdensity factor was $\sim$1.05 and
$\sim$1.16 for GOODS-N and EGS respectively. This first estimation
tell us that we are observing fields with a higher density of galaxies than the mean density in the redshift range 0.75 $<$ z $<$ 0.9. However, at the small range covered by our narrow-band filter densiites could be very different. Fortunately, spectroscopic
redshift surveys are precise enough to reveal the structure in
redshift ranges as small as our. Then, we measured the density ratio
of objects with reliable spectroscopic redshift within our redshift
range over those within 0.75 $<$ z $<$ 0.9. For GOODS-N we found that
this factor was $\sim$1.9, which translates to $\sim$2.0 when we take
into account the density factor for GOODS-N over the mean density. For
Groth2 and Groth3 fields we first measured the density factors between
these fields and the whole EGS field for the redshift range 0.75 $<$ z
$<$ 0.9. We obtained $\sim$1.07 and $\sim$0.71 for Groth2 and Groth3
respectively, showing that could be high variations from field to
field. Then, we measured the ratios between the galaxies within our
redshift range and the galaxies in the wider redshift range, obtaining
$\sim$2.27 and $\sim$1.30, which become $\sim$2.43 and $\sim$0.93 for
Groth2 and Groth3 fields respectively when compared to the whole
EGS. Finally, applying the overdensity factor of the EGS, we
obtained the final factors: $\sim$2.8 and $\sim$1.08 for Groth2 and
Groth3 fields respectively. 

We applied the same method to compute the luminosity function once we
applied the extinction and field to field variance correction.  The
resulting best fit to a Schechter function gives:

\begin{eqnarray*}
\nonumber \phi^*=10^{-2.76\pm0.32} {\rm Mpc}^{-3}\\
\nonumber L^*=10^{42.97\pm0.27} {\rm erg}\ {\rm s}^{-1} \\
\nonumber \alpha=-1.34\pm0.18\\
\end{eqnarray*}

\begin{figure}
\includegraphics[width=8.5cm]{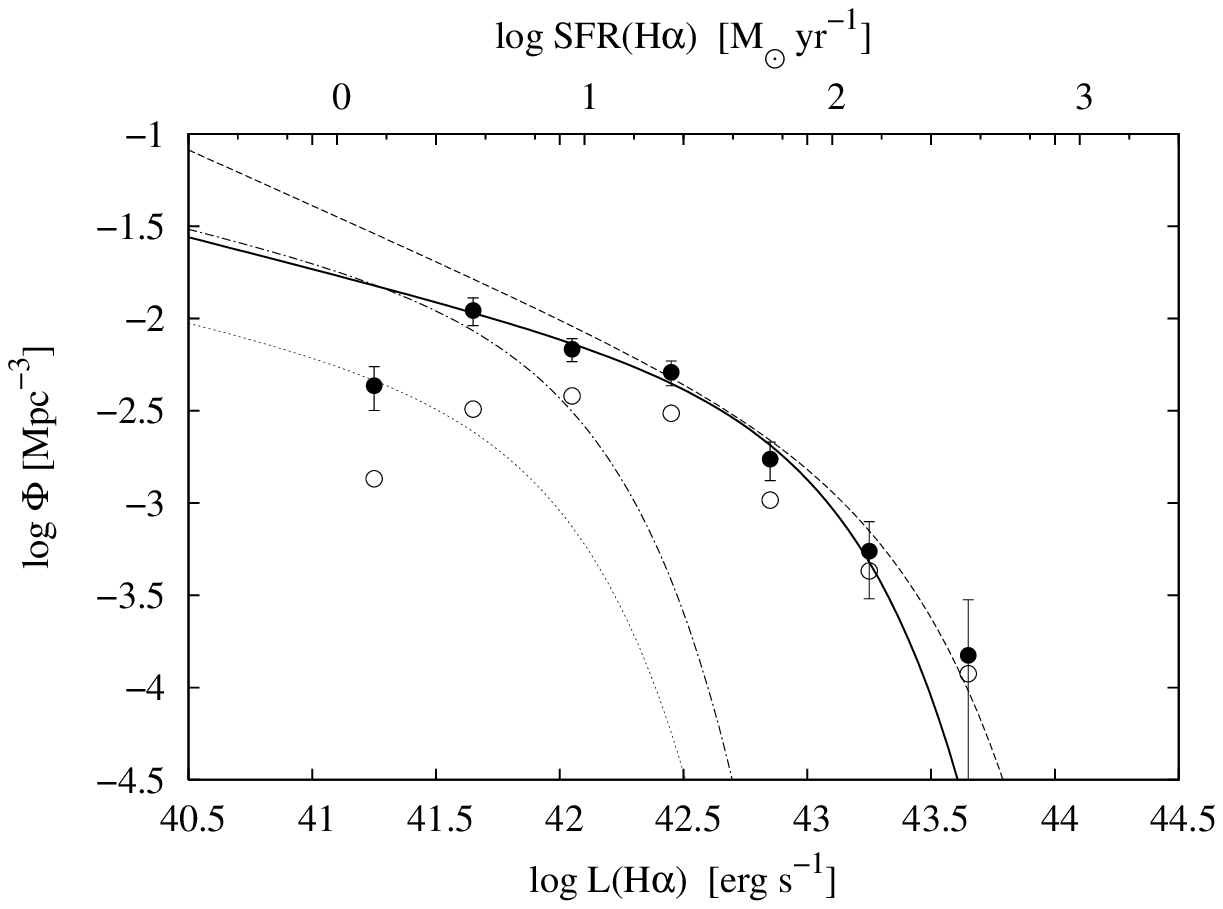}
\caption{\label{lf_red} H$\alpha$ luminosity function corrected 
for extinction (solid circles) with the best fit to a Schechter
function (thick line). Open circles represent the derived LF before
applying the completeness correction. \cite{Gallego95} local LF
(dotted line) is also shown. \cite{Tresse02} LF (dot-dashed line) and
\cite{Hop00} LF (dashed line) corrected for extinction are also
shown. No correction was applied to the \cite{Hop00} LF originally so
we applied the typical A(H$\alpha$)=1 mag. }
\end{figure}

Note that this time we also fitted the faint-end slope. In the fitting
process, we discarded the faintest and the brightest bins. The
faintest bin was clearly affected by incompleteness. The brightest bin
fell off the general shape of the best Schechter fit. Moreover, it
contains only one object, which could be there due to a wrong
estimation of the reddening or a photo-z outlier.  Figure~\ref{lf_red}
shows the extinction-corrected LF derived in this work as well as
\cite{Tresse02} and \cite{Hop00} corrected for extinction
LFs. \cite{Tresse02} applied an overall extinction correction
A$_{V}$=1 mag obtained from the CFRS sample, except for two galaxies
where high quality spectra were available and f(H$\beta$) and
f(H$\delta$) could be measured. \cite{Hop00} did not attempt the
extinction correction though we can apply the typical correction
A(H$\alpha$)=1 mag \citep[see ][ and references therein]{P07} for this
kind of surveys.

The change in the shape of the LF after correcting for extinction and
field to field variance is evident. The typical H$\alpha$ luminosity
has increased more than 1 dex, from $\log$ L$^*$(H$\alpha$)=41.69 to
$\log$ L$^*$(H$\alpha$)=42.97, and the density $\phi^{*}$ has
decreased from -1.74 to -2.76, although this is explained in part because no density correction was applied to the observed LF. Now we can see a clear difference between \cite{Tresse02} LF and this work in the bright regime.

In order to check if the bright regime may be affected by errors in extinction, we repeated the procces to obtain the Schecter parameters, including a typical error in extinction of 0.3 mag. The effect on the Schecter parameteres was found negligible.

Applying an individual
extinction to each object modifies the whole shape of the LF, because
the objects with highest corrected H$\alpha$ luminosities present high
extinctions. Note that most of the previously published H$\alpha$ LFs
assume an average extinction. However, for the total integrated SFRd,
we obtain very similar results with both approaches (see
section~\ref{sfrd}).

Now the shape of our LF is very similar to that of \cite{Hop00},
although they still present a higher density at faint
luminosities. However, we have applied a global extinction correction
to their LF, so we might expect a change in shape and an increase in
luminosity if we make a careful extinction correction. Moreover, we
obtain a mean extinction A(H$\alpha$)=1.48 in our sample and we expect
even higher attenuation as we move to higher redshifts, so probably
their LF would move to higher H$\alpha$ luminosities.

\subsection{The H$\alpha$-based cosmic star formation rate density}
\label{sfrd}

Once we have the LF, we can compute the H$\alpha$ luminosity density through:

\begin{equation}
 \rho _L (H\alpha)=\phi^*\space L^*\space \Gamma (2+\alpha)
\end{equation}

where $\phi ^*$, $L^*$ and $\alpha$ are the parameters obtained in the
Schechter fitting to the LF.

We convert this luminosity density to star formation rate density
through the \cite{Ken98} calibration.

%  Recently \cite{Pfla07} showed
% that such a linear relation is not valid for low H$\alpha$
% luminosities, becoming non-linear below $\sim$10$^{39}{\rm erg}\ {\rm
% s}^{-1}$. This is not a major concern for our results, since our
% luminosities are a few orders of magnitude above that regime.
 We find
that the inferred extinction-, field to field variance-corrected star
formation rate density is $\dot\rho_*$=0.17$^{+0.03}_{-0.03}$
M$_{\odot}$ yr$^{-1}$ Mpc$^{-3}$.

As a consistency check, we checked that the observed and extinction corrected SFR densities differ by the mean extinction correction. We integrated the observed LF and applied the mean extinction correction. However, the observed LF was also affected by field to field variance so we applied a mean density correction (see section~\ref{cor_lumfunc}). The SFRd obtained in this case is $\dot\rho_*$=0.19$^{+0.03}_{-0.03}$ M$_{\odot}$ yr$^{-1}$ Mpc$^{-3}$ , in good agreement with the previous value. This shows that although a mean extinction correction may not be the appropriate method to obtain real shape of the LF, it is enough to accurately determine the luminosity density or SFRd.

This value has not been corrected for AGN contribution as the effects are very small and other authors have not corrected their values either. The AGN contamination is a very difficult to solve problem, and a detailed analysis is out of the scope of this paper. We have tried to quantify how many of our galaxies harbor a luminous AGN by cross-correlating our sample with X-ray catalogs. We looked for X-ray detections in the Chandra 2Ms X-ray point source catalog \citep{Alex03} in GOODS-N. We found 4 X-ray detections out of 58 candidates, within a 2$\arcsec$ search radius. The amount of H$\alpha$ flux concentrated in these sources is 10\% of the total flux in the whole GOODS-N sample, whereas their contribution to the number of galaxies is 8\% (4/58). This result is in good agreement with \cite{Do06} who found an AGN upper limit contribution of 9.5\% to the flux density. \cite{Gallego95} found higher values for the UCM local sample: 10\% in number and 15\% in flux density. In any case, it is important to notice that, although X-ray emission primarily come from the AGN, H$\alpha$ emission could come from a mixture of star forming processes and AGN activity. Hence, the fraction of H$\alpha$ flux concentrated in the X-ray detected sources is an upper limit to the H$\alpha$ flux coming from AGN activity. These X-ray catalogs could be missing very obscured AGN. We have checked the MIR SED of all our objects and none of them would qualify as a power-law galaxy (i.e., a heavily extincted AGN; see, e.g., \citealt{Alo06}). Still, even for the X-ray emitters, it would be impossible to quantify (with the data in our hands) whether the AGN or star formation dominate the H $\alpha$ (or MIR) emission.

\begin{figure}
\includegraphics[width=8.5cm]{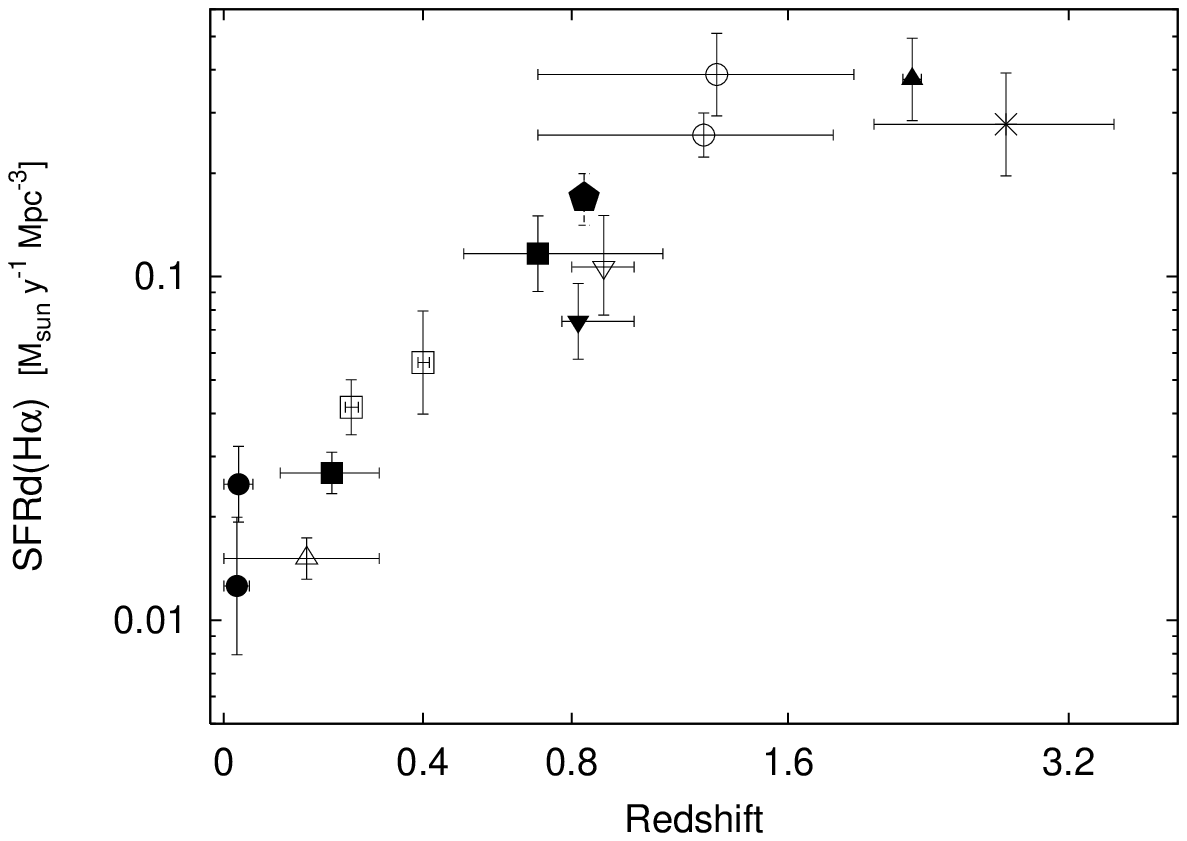}
\caption{\label{sfrd_fig} Evolution of the star formation rate 
density with redshift (scale is given by log (1+z)) for estimations
based on H$\alpha$ measurements. The dark pentagon is this work
result. Other H$\alpha$ measurements come from \cite{Gallego95} and
\cite{PG03} (filled circles), \cite{Sul00} (triangle), \cite{P05}
(empty squares), \cite{Tresse98} and \cite{Tresse02} (filled squares),
\cite{Do06} (inverted filled triangle), \cite{gla99} (inverted
triangle), \cite{Yan99} and \cite{Hop00} (empty circles), \cite{Mor00}
(filled triangle), and \cite{Pet01} (star).}
\end{figure}

We compare our result with other SFRd measured via H$\alpha$ line flux
in figure~\ref{sfrd_fig}. We took the different values from
\cite{Hop04} except values at z=0.24 and z=0.4 that were taken from
\cite{P05} and the result at z$\sim$0.82 by \cite{Do06}. The
\cite{Pet01} value, obtained via H$\beta$, is also shown. In the
\cite{Hop04} compilation, all SFRd values were corrected for
extinction using a SFR dependent obscuration when the LF was available
and the correction by the original authors (if any) was overall and
SFR independent. If the LF was not available and no correction was
made by the original authors, a mean obscuration correction of
A(H$\alpha$)=1 mag was applied.

The closest values in redshift are those by \cite{Tresse02},
\cite{Do06}, and \cite{gla99}. Our result is systematically higher,
about a factor of $\sim$1.5. However in the figure we can see that the
difference between \cite{Tresse02} and this work could be an evolution
effect, as they follow the general trend in redshift (evolution will
be discussed in more detail in Section~\ref{sfrd_evo}). The other two
points fall off the general trend. \cite{Do06} value was corrected for
incompleteness by a factor of $\sim$3 due to the inherent difficulty
of multi-object fiber spectroscopy observations. Only 9 out of 38
galaxies observed were clearly detected ($\geq$5$\sigma$) and the
others were stacked in order to get some information about H$\alpha$
low luminosity objects. They also had to apply aperture corrections,
with a mean value of 2.4 but with some individual values above 4. In
spite of all the efforts they put in to correct for incompleteness and
flux loss, they could still be missing an important fraction of flux
density. Another bias that could have affected their result is that
the selection was made in the {\em R} band, which samples
$\sim$3600\AA\ rest-frame (i.e. the {\em U} band), taking the
precaution to select only targets with identified emission-lines. The
{\em U}-band, although a good tracer of star forming galaxies
\citep{Mous06}, is not a direct tracer as can be the rest-frame UV. In
addition, as pointed out in Section~\ref{ha_lf}, 20\% of objects with
no [O{\sc ii}]$\lambda$3727 emission have H$\alpha$ emission in the
SDSS, so they could be missing a fraction of H$\alpha$
emitters. \cite{gla99} SFRd is $\sim$2 times lower than our
value. However they only detected 8 galaxies with H$\alpha$ in
emission, which could severely affect their results.

\subsection{Star formation rate density evolution}
\label{sfrd_evo}
It is obvious from figure~\ref{sfrd_fig} that a decrease in SFRd has
occurred from z$\sim$1 to the local Universe, being the latter
$\sim$10 times less active forming stars. It is common to parametrize
the evolution of the SFR density with a power law:
$\dot\rho_{*}\propto (1+z)^{\beta}$.

\begin{figure}
\includegraphics[width=9cm]{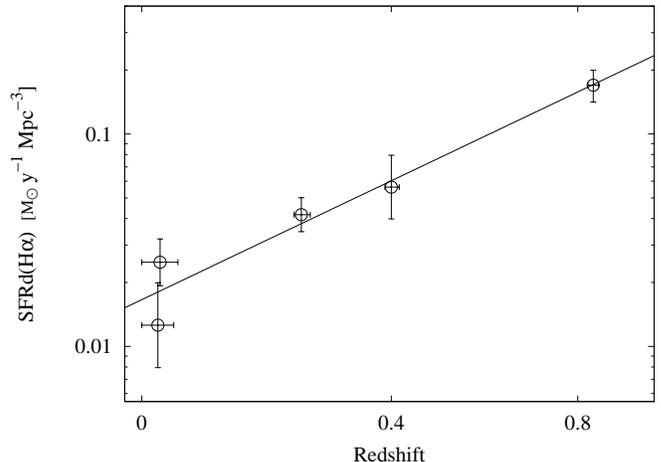}
\caption{\label{sfrd_evo_fig} Evolution of SFRd measured by the UCM 
through the H$\alpha$ line (\citealp{Gallego95}, \citealp{PG03},
\citealp{P05} and this work). The line represents the best fit to a
law $\propto$(1+z)$^\beta$ with $\beta$=3.8.}
\end{figure}

Combining just the H$\alpha$-based SFR densities obtained by our group
at $z=0.02$, $z=0.24$, $z=0.40$ and $z=0.84$ for H$\alpha$-selected
samples, we obtain an evolution of the cosmic SFR density
$\propto(1+z)^\beta$ where $\beta=3.8\pm0.5$. The fitted power law is
shown in figure~\ref{sfrd_evo_fig}. This $\beta$ value is similar to
the one estimated by \cite{Tresse02} using H$\alpha$ observations, and
by
\cite{PG05} for a thermal IR-selected sample. However, \cite{Tresse02} 
value was calculated for an Einstein-de Sitter cosmology with H$_{\rm
0}$=50 km$^{-1}$. The cosmology change softens this value to $\approx$
3.5 \citep{Do06}. Thus, we find a slightly higher value but still
compatible within errors.  Our value is also comparable to that of
\cite{Hop04} who used data obtained with multiple star formation
tracers and obtained $\beta$=3.19$\pm$0.26 for a luminosity dependent
obscuration correction.

\begin{figure}
\includegraphics[width=8.5cm]{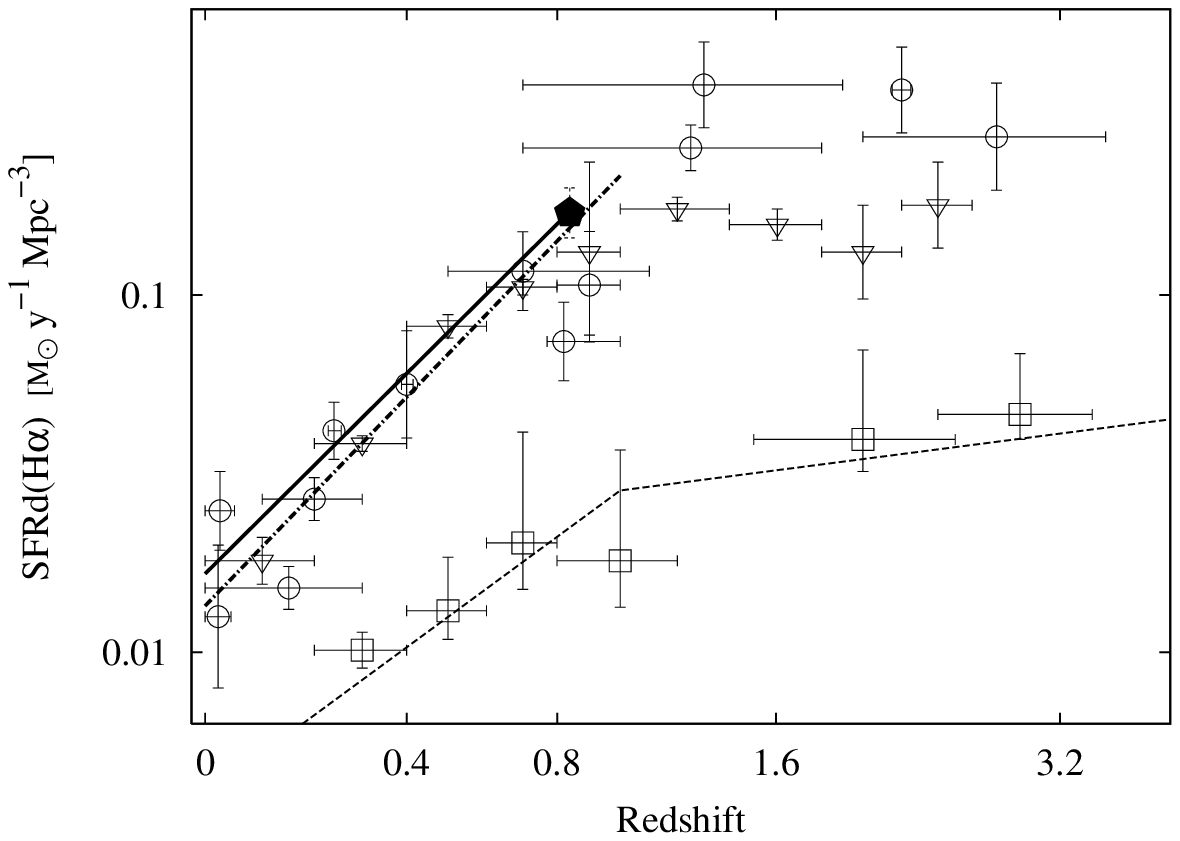}
\caption{\label{sfrd_evo_mw_fig} Evolution of the star formation rate 
density with redshift for estimations based on H$\alpha$, IR and FUV
measurements. The dark pentagon is this work result. Other H$\alpha$
measurements (open circles) come from \cite{Gallego95}, \cite{PG03},
\cite{Sul00}, \cite{P05}, \cite{Tresse98}, \cite{Tresse02},
\cite{Do06}, \cite{gla99}, \cite{Yan99}, \cite{Hop00}, \cite{Mor00}
and \cite{Pet01}. IR measurements \citep{PG05} are represented as open
triangles and FUV estimations \citep{Schimi05} as open squares. The
derived evolution law is represented as a thick line for H$\alpha$, a
dashed line for UV and a dot-dashed line for IR. }
\end{figure}

More interesting is to compare the evolution of the SFRd obtained
through different estimators. In Figure~\ref{sfrd_evo_mw_fig} we plot
the SFRd history obtained in H$\alpha$, IR \citep[from ][]{PG05} and
UV \citep[from ][]{Schimi05}.  H$\alpha$-based SFRd values are
corrected for reddening while the UV values are not. \cite{PG05}
obtained $\beta$=3.98$\pm$0.22 for a IR selected sample and
\cite{Schimi05} obtained $\beta$=2.5$\pm$0.7, both up to z=1. Our
H$\alpha$ measurement agrees quite well with that obtained with the IR
sample. The UV slope is significantly lower, which could be caused by
an evolution in the extinction properties with redshift, in the case
the populations selected with each method were mainly the same.

There is another interesting question that arises when comparing
populations selected with different observables: are we selecting the
same objects or are there substantial differences?

To answer this question, we have considered the galaxies detected by
MIPS and having a reliable spectroscopic redshift. There are 11, 2 and
18 objects in Groth2, Groth3 and GOODS-N respectively that are within
our filter redshift range, are detected by MIPS, but not selected in
our survey. This imply a 17\%, 8\% and 47\% of the total 24$\mu$m flux
in spectroscopically confirmed MIPS galaxies (at z$\sim$0.84),
corresponding the lowest fraction to our deepest field and the high
fraction to our shallowest filed. Thus, the extinction in these
objects could make their H$\alpha$ flux fall below our detection
limits, being worst for the shallower fields. However, the fraction in
GOODS-N is still quite high to be explained by the different field
depths. The explanation comes from the different extinction for the
objects in these fields. Whereas for the Groth fields we have a mean
$\overline{A}(H\alpha)$=1.75 mag, somewhat higher than the mean value
for the whole sample, for the GOODS-N field this value is
$\overline{A}(H\alpha)$=3 mag.

On the other hand, we are missing 11, 2 and 22 objects detected in the
GALEX NUV-band which is very close to rest-frame FUV (a good estimator
of the SFR). Most of these objects are missed because they fall below
our detection limit. GALEX reach smaller SFRs, but only in the case of
low attenuation. However, some objects show UV emission corresponding
to a star formation rate that could be selected by our method. These
missed objects could also be post starburst (that over predict current
star formation) although there is no H$\alpha$ emission.

The opposite case is also present. We detect 23 (41\%), 32 (63\%) and
20 (34\%) objects (including those with only photometric redshift)
that do not show MIPS 24$\mu$m emission. The mean star formation rate
for these objects is 2.8, 2.2 and 2.6 M$_{\odot}$ yr$^{-1}$ with mean
extinctions of H$\alpha$ of 1.0, 0.9 and 1.3 mag, thus having a mean
corrected star formation rate of 7.0, 5.0 and 8.6 M$_{\odot}$
yr$^{-1}$. These values are below the 80\% completeness limit of the
MIPS instrument in these fields: 83 $\mu$Jy which corresponds to
$\sim$10 M$_{\odot}$ yr$^{-1}$ \citep{PG05}. When considering the UV
emission there are 37(66\%), 28(55\%) and 26(44\%) objects not
detected in the GALEX NUV band. The mean H$\alpha$ star formation rate
for these objects is 2.7, 2.5 and 3.45 M$_{\odot}$ yr$^{-1}$ with the
following mean extinctions: 1.5, 1.5 and 1.6 magnitudes in
H$\alpha$. If we translate these SFRs to observed SFRs in the UV we
obtain 1.4, 1.34 and 1.7 M$_{\odot}$ yr$^{-1}$. These low SFRs are
similar to the detection limit for the GALEX NUV-band ($\sim$1.5
M$_{\odot}$ yr$^{-1}$) at z=0.84 based on the analysis of the GALEX
catalog.

As we have shown, the most significant loss is the FIR emitters
because they have high SFRs, but lie below our detection limit due to
the presence of dust. We notice that although we are losing a fraction
of the FIR objects our completeness correction is also recovering a
fraction of them. The UV objects not recovered in our sample are very
faint and contribute to the low luminosity regime of the luminosity
function. On the other hand, FIR and UV surveys miss a significant
fraction of objects. In the case of FIR it is worth to notice that,
although we are missing a fraction of star forming galaxies, we obtain
a very similar SFRd value to that of \cite{PG05} (even higher). Hence, the
objects not detected by MIPS with lower SFRs are playing a more
important role than that estimated by \cite{PG05}. We conclude that
our work is complementary to FIR and UV surveys, as it goes fainter
than FIR detection limits and is not as affected by extinction as the
UV.

\section{Summary and conclusions}
Using an H$\alpha$ selected sample of star-forming galaxies we have
estimated the H$\alpha$ luminosity function for the Universe at
$z=0.84$. This work is the continuation of previous surveys where our
group used the H$\alpha$ emission to select representative samples of
star-forming galaxies at intermediate redshifts.  We argue that, since
the H$\alpha$ emission provides a good estimate of the instantaneous
star formation, the galaxies have been selected in a homogeneous way
up to z$\sim$1 by their current SFR. Therefore, we can use the
H$\alpha$ luminosity function to determine the ``current SFR
function'' describing the number of star-forming galaxies as a
function of their SFR. Integrating over all H$\alpha$ luminosities (or
SFRs) we determine the current SFR density for galaxies.

A total of 165 objects have been selected as H$\alpha$ emitters using
the narrow band technique. We have tested the reliability of our
emission-line candidates in three different ways: 1) analyzing the use
of photometric apertures of different sizes; 2) carrying out a
star-galaxy segregation; and 3) estimating photometric redshifts.

Line luminosities have been corrected for Nitrogen contribution and
dust reddening. To correct for Nitrogen flux contamination, we used
the SDSS sample to estimate the mean [NII] to H$\alpha$ flux ratio for
a given EW(H$\alpha$+[NII]). For the dust reddening correction, we
proceeded in several steps: for the objects with MIPS detection, we
estimated the total infrared luminosity and a synthetic FUV flux from
templates and then compute the IRX ratio, which is related to the
extinction in the FUV band. If the object was not detected in MIPS we
used the UV slope, given by FUV-NUV \citep{Gil06}. We found a mean
extinction for the whole sample A(H$\alpha$)=1.45 mag, ranging from 0
to 4.16.

We performed simulations to determine the limiting flux and
completeness corrections. The limiting fluxes vary from field to field
from 8$\times10^{-17}$ erg\,s$^{-1}$\,cm$^{-2}$ to 14 $\times
10^{-17}$ erg\,s$^{-1}$\,cm$^{-2}$ for a 70\% completeness level. The
completeness correction was applied to the computation of the
extinction not corrected and corrected LF.

We computed the observed LF not corrected for extinction, obtaining
the following parameters when fitting to a Schechter function:
$\phi^*$=$10^{-1.74\pm0.11}$ Mpc$^{-3}$, L$^*=10^{41.69\pm0.07}$ erg
s$^{-1}$. We fixed the low-luminosity slope $\alpha$=-1.35 because our
LF was not deep enough. Our LF has higher density than that of
\cite{Tresse02}, that could be explained by an evolutionary effect due
to the different mean redshifts explored or by the selection method in
each case: directly by H$\alpha$ in this work whereas they had to use
the $I$-band and spectroscopic redshifts. \cite{Yan99} and
\cite{Hop00} LFs extend to higher luminosities than ours. A possible
explanation could be that these authors surveyed a higher redshift
range (up to z$\sim$1.9), where star forming galaxies properties could
be significantly different than at z$\sim$0.84.

The LF corrected for extinction and field to field variance yielded:
$\alpha$=-1.34$\pm$0.18, $\phi^*$=10$^{-2.76\pm0.32}$ Mpc$^{-3}$ and
$L^*$=10$^{42.97\pm0.27}$ erg s$^{-1}$. The LF extends now to
similar luminosities than \cite{Hop00} LF, although, as in the
original work no extinction correction was applied, we applied a mean
correction A(H$\alpha$)=1. However, this mean correction could lead to
subestimate L$^*$, as the highest attenuated sources wouldn't move to
their actual high luminosities. On the other hand, we found a mean
attenuation for our sample A(H$\alpha$)=1.45 whereas we have applied
the typical mean correction for the \cite{Hop00} LF.

Analyzing each field independently and compared to the mean density of
galaxies we found that there is an overabundance factor of $\sim$2.8,
$\sim$1.08 and $\sim$2.0 for Groth2, Groth3 and GOODS-N fields
respectively.

The SFRd derived from the extinction and field to field variance
corrected LF is $\dot\rho_{*}$=0.17$^{+0.03}_{-0.03}$ M$_{\odot}$
yr$^{-1}$ Mpc$^{-3}$. The strong increase from z=0. to z$\sim$1 found
in other surveys is confirmed. Combining just the H$\alpha$-based SFR
densities obtained by our group from $z=0.02$, $z=0.24$, $z=0.40$ and
$z=0.84$ H$\alpha$-selected samples, we obtain an evolution of the
cosmic SFR density $\propto(1+z)^\beta$ where $\beta=3.8\pm0.5$. This
$\beta$ value is similar to the one estimated by \cite{Tresse02} and
by \cite{PG05} for thermal IR-selected samples.

The H$\alpha$ approach is complementary to FIR and UV surveys as it
reach fainter SFRs than FIR surveys and is less affected by extinction
than UV surveys. The fraction of objects detected in FIR not detected
by H$\alpha$ is around $\sim$15\% unless very high extincted objects
are present, as in the case of GOODS-N.

\acknowledgments
We thank Nicolas Cardiel and Armando Gil de Paz for advice and useful comments.
V.\ Villar acknowledges the receipt of a \emph{Formaci\'on de
Personal Investigador} fellowship from
the Spanish Ministerio de Educaci\'on y Ciencia.
This work was supported in part by the Spanish \emph{Plan Nacional
de Astronom\'{\i}a y Astrof\'{\i}sica} under grant AYA2003-01676 and AYA2006-02358.
This work was supported in part by a generous grant of the \emph{New del Amo
Foundation Program}.

{\it Facilities:} \facility{CAO:3.5m} \facility{Keck:II}
\bibliographystyle{apj}
\bibliography{referencias_paper,apj-jour}

\clearpage
\setcounter{table}{2}
\begin{deluxetable}{ccccrrcrr}
\tabletypesize{\scriptsize}
\tablecaption{Observed sample}
\tablehead{
\colhead{ID} & \colhead{$\alpha$ (J2000)} & \colhead{$\delta$ (J2000)} & \colhead{J}& \colhead{f$_{H_{\alpha}}$} & \colhead{EW(H$\alpha$)} & \colhead{A(H$\alpha$)} & \colhead{SFR$_{obs}$} & \colhead{SFR$_{cor}$} \\
\colhead{(1)} & \colhead{(2)} & \colhead{(3)} & \colhead{(4)}& \colhead{(5)} & \colhead{(6)} & \colhead{(7)} & \colhead{(8)}& \colhead{(9)}
}
\startdata
f2\_481&214.39872&52.36067&20.22$\pm$0.08&23.6$\pm$4.7&67&0.45&4.6&6.9\\
f2\_566&214.49003&52.36594&21.62$\pm$0.12&15.4$\pm$3.9&209&1.81&3.2&16.9\\
f2\_592&214.28326&52.36750&21.88$\pm$0.16&14.1$\pm$4.4&241&1.60&3.0&13.0\\
f2\_688&214.26028&52.37098&21.76$\pm$0.14&14.1$\pm$4.4&121&3.04&2.9&48.4\\
f2\_959&214.44781&52.37737&19.15$\pm$0.03&6.8$\pm$3.4&17&1.97&1.2&7.4\\
f2\_997&214.43842&52.38060&21.20$\pm$0.09&14.2$\pm$4.0&132&1.04&2.8&7.3\\
f2\_1238&214.40850&52.38895&19.95$\pm$0.05&20.9$\pm$5.6&34&2.02&3.9&24.9\\
f2\_1461&214.21682&52.39630&18.95$\pm$0.02&53.8$\pm$7.8&33&2.08&10.0&68.1\\
f2\_1463&214.40130&52.39658&22.26$\pm$0.27&11.1$\pm$4.1&282&1.06&2.4&6.4\\
f2\_1650&214.49389&52.40403&21.12$\pm$0.09&25.6$\pm$6.0&111&1.25&5.3&16.6\\
f2\_1694&214.36583&52.40625&21.12$\pm$0.12&14.8$\pm$3.8&179&1.41&3.0&11.0\\
f2\_1822&214.20626&52.41180&22.48$\pm$0.25&9.2$\pm$3.7&273&0.07&2.0&2.1\\
f2\_2091&214.50282&52.41996&21.82$\pm$0.19&30.1$\pm$4.3&513&0.08&7.2&7.7\\
f2\_2121&214.51324&52.42082&21.70$\pm$0.17&12.8$\pm$4.1&231&1.65&2.7&12.2\\
f2\_2219&214.25459&52.42487&21.40$\pm$0.11&20.0$\pm$4.6&115&0.72&4.1&8.0\\
f2\_2362&214.50548&52.42970&21.57$\pm$0.12&16.0$\pm$4.2&114&1.12&3.3&9.2\\
f2\_2522&214.46666&52.43515&23.48$\pm$0.32&10.9$\pm$3.2&611&0.23&2.7&3.3\\
f2\_2539&214.50109&52.43483&20.83$\pm$0.08&8.1$\pm$4.3&32&3.33&1.5&32.4\\
f2\_2556&214.30817&52.43651&22.47$\pm$0.21&17.1$\pm$4.3&450&0.07&4.0&4.3\\
f2\_2598&214.49855&52.43669&20.71$\pm$0.07&18.7$\pm$5.6&59&0.92&3.6&8.3\\
f2\_2668&214.31872&52.43925&20.84$\pm$0.13&22.4$\pm$5.1&189&1.34&4.6&15.6\\
f2\_2706&214.50345&52.44056&20.67$\pm$0.09&40.4$\pm$7.0&223&1.04&8.4&22.0\\
f2\_2993&214.23302&52.44979&19.86$\pm$0.06&20.6$\pm$5.8&35&1.75&3.8&19.3\\
f2\_3093&214.24402&52.45767&22.55$\pm$0.29&9.3$\pm$3.7&275&0.45&2.0&3.0\\
f2\_3166&214.21455&52.45995&21.72$\pm$0.21&16.0$\pm$5.2&292&0.84&3.5&7.6\\
f2\_3194&214.24633&52.45995&21.16$\pm$0.12&10.0$\pm$2.9&74&2.46&2.0&18.9\\
f2\_3345&214.25875&52.46516&20.67$\pm$0.07&16.1$\pm$4.2&52&2.75&3.1&38.7\\
f2\_3358&214.24843&52.46479&20.64$\pm$0.09&22.4$\pm$5.4&74&2.31&4.4&36.9\\
f2\_3458&214.20323&52.46636&19.19$\pm$0.03&27.5$\pm$6.6&24&1.47&5.0&19.5\\
f2\_3460&214.53003&52.46774&19.95$\pm$0.05&18.8$\pm$5.2&37&1.43&3.5&13.1\\
f2\_3507&214.21475&52.47060&19.48$\pm$0.04&20.0$\pm$5.1&26&1.98&3.7&22.8\\
f2\_3614&214.41869&52.47512&22.75$\pm$0.26&7.7$\pm$2.5&342&2.24&1.7&13.7\\
f2\_3726&214.36825&52.47954&22.64$\pm$0.33&8.3$\pm$3.1&340&0.34&1.9&2.6\\
f2\_3742&214.32370&52.47898&21.12$\pm$0.11&10.4$\pm$3.4&71&2.54&2.0&20.9\\
f2\_3786&214.42755&52.47950&19.44$\pm$0.04&60.5$\pm$7.4&65&2.16&11.7&86.1\\
f2\_3992&214.28128&52.48944&20.03$\pm$0.05&25.8$\pm$6.3&43&1.27&4.9&15.6\\
f2\_4054&214.19562&52.49363&22.31$\pm$0.22&8.9$\pm$3.2&287&0.04&1.9&2.0\\
f2\_4179&214.45653&52.49805&23.14$\pm$0.32&12.0$\pm$3.9&549&1.41&2.9&10.6\\
f2\_4323&214.25305&52.50370&21.13$\pm$0.13&9.7$\pm$3.4&77&2.34&1.9&16.5\\
f2\_4422&214.53702&52.50786&23.18$\pm$0.32&20.1$\pm$3.4&1077&0.38&5.1&7.2\\
f2\_4692&214.23025&52.51940&21.68$\pm$0.14&6.1$\pm$2.8&129&1.02&1.2&3.0\\
f2\_4713&214.29729&52.51900&20.15$\pm$0.07&25.5$\pm$5.2&63&0.85&4.9&10.8\\
f2\_4769&214.28798&52.52265&23.26$\pm$0.28&8.9$\pm$3.2&505&3.74&2.1&65.8\\
f2\_4937&214.30355&52.52710&19.77$\pm$0.05&12.6$\pm$4.1&27&2.76&2.3&29.2\\
f2\_4986&214.43021&52.52997&22.48$\pm$0.21&11.4$\pm$3.5&190&1.16&2.6&7.5\\
f2\_5405&214.53927&52.54503&20.83$\pm$0.10&18.1$\pm$5.6&69&1.12&3.5&9.9\\
f2\_5408&214.53529&52.54495&20.28$\pm$0.06&79.5$\pm$6.7&167&1.02&17.5&45.1\\
f2\_5459&214.47510&52.54793&22.12$\pm$0.18&7.7$\pm$4.7&150&0.75&1.5&3.1\\
f2\_5508&214.28716&52.56664&21.12$\pm$0.10&35.9$\pm$6.1&154&0.63&7.8&14.0\\
f2\_5584&214.30300&52.56433&21.20$\pm$0.10&20.0$\pm$5.2&95&1.26&4.0&12.8\\
f2\_5594&214.55351&52.56617&22.04$\pm$0.25&31.9$\pm$8.8&646&0.44&7.8&11.7\\
f2\_5639&214.28714&52.56724&21.16$\pm$0.10&11.2$\pm$3.8&68&1.04&2.2&5.6\\
f2\_5959&214.29914&52.55195&21.79$\pm$0.20&13.8$\pm$4.2&136&3.04&2.9&48.2\\
f2\_5993&214.55324&52.54776&19.99$\pm$0.06&18.7$\pm$9.8&42&2.88&3.5&50.0\\
f2\_7462&214.28432&52.56822&20.50$\pm$0.06&27.1$\pm$5.0&77&1.26&5.3&17.1\\
f3\_530&214.35434&52.58357&21.47$\pm$0.12&27.0$\pm$6.2&288&0.68&5.9&11.0\\
f3\_578&214.54496&52.58483&22.35$\pm$0.18&12.0$\pm$4.3&280&1.31&2.6&8.7\\
f3\_863&214.60952&52.58966&19.46$\pm$0.03&17.5$\pm$4.8&20&1.34&3.2&10.9\\
f3\_1282&214.66531&52.60264&19.19$\pm$0.02&29.3$\pm$6.3&25&1.10&5.3&14.8\\
f3\_1316&214.51123&52.60716&21.41$\pm$0.09&16.3$\pm$3.9&94&4.13&3.3&147.8\\
f3\_1344&214.69324&52.60773&23.53$\pm$0.31&11.1$\pm$3.3&599&1.25&2.7&8.6\\
f3\_1390&214.37272&52.60940&21.53$\pm$0.10&15.2$\pm$3.7&99&1.93&3.1&18.1\\
f3\_2440&214.71246&52.64082&20.89$\pm$0.07&13.8$\pm$4.0&98&1.04&2.6&6.9\\
f3\_2588&214.40606&52.64745&21.81$\pm$0.12&10.2$\pm$3.6&158&1.06&2.0&5.4\\
f3\_2634&214.65503&52.64757&20.41$\pm$0.05&23.0$\pm$5.4&98&2.02&4.4&28.2\\
f3\_2635&214.70055&52.64851&22.29$\pm$0.14&10.5$\pm$3.2&245&1.00&2.2&5.6\\
f3\_2694&214.40516&52.65156&21.87$\pm$0.17&9.4$\pm$2.8&246&0.58&2.0&3.4\\
f3\_2919&214.66029&52.65804&21.60$\pm$0.12&7.4$\pm$2.7&123&1.14&1.4&4.1\\
f3\_3041&214.75493&52.66071&21.29$\pm$0.11&10.9$\pm$4.6&103&0.08&2.1&2.2\\
f3\_3112&214.42338&52.66452&21.34$\pm$0.12&4.7$\pm$2.2&111&0.92&0.9&2.1\\
f3\_3417&214.57087&52.67417&20.32$\pm$0.04&16.2$\pm$3.7&41&1.86&3.0&16.9\\
f3\_3525&214.41201&52.67866&21.26$\pm$0.10&8.6$\pm$2.9&136&1.06&1.7&4.5\\
\enddata
\tablecomments{(1) Identification (2) RA (J2000) (3) DEC (J2000) (4) J-magnitude (Vega) (5) Line flux (10$^{-17} erg s^{-1} cm^{-2}$ (6) Restframe equivalent width (\AA) (7) Extinction in magnitudes (8) SFR not corrected for extinction (M$_{\odot}$ yr$^{-1}$) (9) SFR corrected for extinction (M$_{\odot}$ yr$^{-1}$).}
\label{tab_objects}
\end{deluxetable}

\setcounter{table}{2}
\begin{deluxetable}{ccccrrcrr}
\tabletypesize{\scriptsize}
\tablecaption{Observed sample}
\tablehead{
\colhead{ID} & \colhead{$\alpha$ (J2000)} & \colhead{$\delta$ (J2000)} & \colhead{J}& \colhead{f$_{H_{\alpha}}$} & \colhead{EW(H$\alpha$)} & \colhead{A(H$\alpha$)} & \colhead{SFR$_{obs}$} & \colhead{SFR$_{cor}$} \\
\colhead{(1)} & \colhead{(2)} & \colhead{(3)} & \colhead{(4)}& \colhead{(5)} & \colhead{(6)} & \colhead{(7)} & \colhead{(8)}& \colhead{(9)}
}
\startdata

f3\_3675&214.42497&52.68249&19.15$\pm$0.02&28.2$\pm$6.7&22&1.29&5.1&16.7\\
f3\_3694&214.40237&52.68428&20.77$\pm$0.06&20.6$\pm$4.4&69&0.66&4.0&7.3\\
f3\_3972&214.54982&52.69515&21.85$\pm$0.15&11.8$\pm$3.6&186&0.08&2.4&2.6\\
f3\_4116&214.42933&52.70046&22.28$\pm$0.18&8.3$\pm$2.8&184&1.54&1.7&6.9\\
f3\_4119&214.75000&52.69880&20.75$\pm$0.07&12.6$\pm$4.3&79&2.20&2.4&18.1\\
f3\_4133&214.56011&52.70091&22.28$\pm$0.17&9.9$\pm$3.1&248&0.08&2.1&2.3\\
f3\_4222&214.48923&52.70217&19.91$\pm$0.04&22.2$\pm$5.6&36&0.52&4.2&6.7\\
f3\_4368&214.42725&52.70791&20.10$\pm$0.04&21.4$\pm$4.2&46&3.75&4.0&128.2\\
f3\_4659&214.34934&52.71719&20.37$\pm$0.06&6.5$\pm$6.3&15&4.04&1.1&47.0\\
f3\_4741&214.41970&52.72037&22.27$\pm$0.15&20.5$\pm$4.2&468&0.08&4.8&5.2\\
f3\_4858&214.37258&52.72191&19.58$\pm$0.03&24.2$\pm$5.1&30&4.17&4.5&207.6\\
f3\_4891&214.40727&52.72368&19.53$\pm$0.02&25.2$\pm$5.1&50&1.25&4.6&14.6\\
f3\_4901&214.54208&52.72290&20.72$\pm$0.08&6.9$\pm$3.0&27&1.96&1.3&7.7\\
f3\_4916&214.63473&52.72382&20.13$\pm$0.04&13.1$\pm$3.9&27&2.38&2.4&21.5\\
f3\_4927&214.36864&52.72580&21.69$\pm$0.12&9.1$\pm$3.3&163&0.07&1.8&1.9\\
f3\_4955&214.40295&52.72560&19.63$\pm$0.03&20.5$\pm$5.2&25&3.58&3.8&101.4\\
f3\_5080&214.60435&52.73046&20.65$\pm$0.08&8.6$\pm$3.0&79&0.04&1.6&1.7\\
f3\_5362&214.71217&52.74022&21.54$\pm$0.16&11.6$\pm$4.0&185&1.31&2.4&7.9\\
f3\_5603&214.56435&52.74962&22.61$\pm$0.28&6.7$\pm$2.5&319&0.95&1.5&3.6\\
f3\_5746&214.75617&52.75273&19.48$\pm$0.03&30.6$\pm$6.4&30&1.65&5.7&25.9\\
f3\_5781&214.64338&52.75524&21.76$\pm$0.12&14.7$\pm$4.0&117&1.34&3.1&10.5\\
f3\_5785&214.50875&52.75358&18.97$\pm$0.02&7.3$\pm$3.2&17&2.33&1.3&11.0\\
f3\_5808&214.48128&52.75600&21.25$\pm$0.09&11.7$\pm$2.9&159&0.84&2.3&5.0\\
f3\_5857&214.67298&52.75791&21.91$\pm$0.13&6.0$\pm$2.4&127&1.00&1.2&2.9\\
f3\_5893&214.75849&52.75765&20.74$\pm$0.08&6.1$\pm$2.7&50&1.14&1.2&3.3\\
f3\_6178&214.38307&52.77068&22.23$\pm$0.14&12.1$\pm$3.4&288&0.64&2.6&4.7\\
f3\_6456&214.36018&52.82435&22.15$\pm$0.15&10.2$\pm$3.7&244&0.38&2.2&3.1\\
f3\_6483&214.39506&52.82217&21.80$\pm$0.19&4.1$\pm$2.2&146&1.41&0.8&3.0\\
f3\_6553&214.46315&52.82027&22.48$\pm$0.25&8.7$\pm$3.2&285&1.25&1.9&6.0\\
f3\_7108&214.44095&52.80083&21.42$\pm$0.13&6.4$\pm$2.5&127&0.04&1.2&1.3\\
f3\_7402&214.51934&52.79224&20.14$\pm$0.06&4.8$\pm$2.2&89&2.06&0.9&6.1\\
f3\_7493&214.47743&52.79039&20.76$\pm$0.06&7.4$\pm$4.1&46&0.30&1.4&1.8\\
f3\_7702&214.39626&52.78315&19.75$\pm$0.04&33.5$\pm$6.4&83&1.65&6.3&29.0\\
g1\_375&189.12078&62.08836&21.86$\pm$0.30&6.8$\pm$3.5&67&1.12&1.3&3.7\\
g1\_456&189.24120&62.09161&22.12$\pm$0.29&5.2$\pm$2.4&196&1.19&1.1&3.2\\
g1\_465&189.19594&62.09143&20.99$\pm$0.12&15.5$\pm$4.1&130&1.95&3.0&18.1\\
g1\_496&188.92963&62.09167&19.12$\pm$0.05&38.9$\pm$9.1&29&2.05&7.2&47.6\\
g1\_516&189.42234&62.09258&20.77$\pm$0.12&12.8$\pm$4.7&54&1.33&2.4&8.3\\
g1\_713&189.22507&62.10214&20.36$\pm$0.12&12.4$\pm$3.2&47&2.65&2.3&26.9\\
g1\_908&189.21466&62.11220&19.97$\pm$0.08&35.5$\pm$5.1&115&2.31&6.8&57.2\\
g1\_1034&189.30751&62.11791&20.88$\pm$0.12&15.0$\pm$3.5&66&0.58&2.9&5.0\\
g1\_1082&189.14916&62.12032&21.44$\pm$0.26&6.8$\pm$2.4&124&1.37&1.3&4.7\\
g1\_1159&189.15771&62.12317&20.66$\pm$0.12&14.6$\pm$3.1&64&1.27&2.8&9.1\\
g1\_1516&189.00670&62.13943&20.59$\pm$0.13&14.1$\pm$3.9&50&2.01&2.7&17.0\\
g1\_1665&189.12674&62.14533&20.87$\pm$0.13&21.4$\pm$3.9&86&1.58&4.2&18.2\\
g1\_1735&189.12719&62.14752&20.34$\pm$0.11&27.2$\pm$4.8&79&3.31&5.4&113.1\\
g1\_1995&189.06608&62.15986&20.93$\pm$0.13&23.8$\pm$4.6&178&0.79&4.8&9.9\\
g1\_2073&189.39735&62.16152&20.71$\pm$0.12&13.4$\pm$4.3&66&3.53&2.6&67.0\\
g1\_2141&189.16036&62.16473&18.85$\pm$0.04&39.5$\pm$6.6&22&1.14&7.2&20.5\\
g1\_2198&189.11342&62.16728&20.39$\pm$0.12&18.2$\pm$3.9&105&0.48&3.5&5.4\\
g1\_2205&189.35772&62.16781&20.94$\pm$0.15&17.2$\pm$5.0&171&1.88&3.4&19.5\\
g1\_2339&189.23896&62.17391&20.90$\pm$0.16&5.9$\pm$2.1&78&1.43&1.1&4.1\\
g1\_2387&188.95030&62.17764&22.01$\pm$0.21&12.0$\pm$3.2&233&0.79&2.5&5.2\\
g1\_2450&188.94491&62.18024&20.56$\pm$0.13&12.1$\pm$3.6&43&2.75&2.3&28.5\\
g1\_2537&188.93729&62.18332&21.46$\pm$0.17&16.2$\pm$3.3&134&1.74&3.4&17.0\\
g1\_2669&189.00505&62.19133&21.77$\pm$0.20&10.1$\pm$2.8&170&2.36&2.0&17.7\\
g1\_2815&189.15753&62.19707&20.64$\pm$0.12&12.8$\pm$4.0&85&2.29&2.4&20.0\\
g1\_2827&189.15960&62.19750&20.44$\pm$0.11&21.0$\pm$4.6&54&3.05&4.0&66.4\\
g1\_2882&189.01208&62.20036&20.63$\pm$0.11&17.9$\pm$4.0&58&0.87&3.4&7.7\\
g1\_3000&189.42950&62.20486&20.38$\pm$0.10&15.7$\pm$8.3&66&1.46&2.9&11.2\\
g1\_3068&188.97302&62.20893&20.06$\pm$0.09&21.7$\pm$4.6&44&0.98&4.1&10.1\\
g1\_3178&189.16642&62.21390&19.55$\pm$0.06&17.8$\pm$3.5&30&2.41&3.3&30.3\\
g1\_3205&189.28483&62.21460&20.72$\pm$0.14&18.0$\pm$4.6&59&1.56&3.5&14.6\\
g1\_3339&189.13999&62.22222&19.91$\pm$0.07&19.3$\pm$4.0&32&2.25&3.6&28.5\\
g1\_3400&189.30463&62.22606&20.96$\pm$0.18&20.0$\pm$3.8&181&0.79&4.0&8.3\\
g1\_3531&188.95300&62.23167&20.57$\pm$0.12&23.4$\pm$3.7&164&0.34&4.7&6.4\\
g1\_3581&189.39394&62.23232&19.00$\pm$0.05&33.9$\pm$10.0&26&1.88&6.2&35.2\\
g1\_3630 & 189.40551 & 62.23646 & 20.31$\pm$0.12 & 9.8  $\pm$  6.3 & 27 & 1.06 & 1.8 & 4.8\\
g1\_3655&189.27626&62.25499&20.23$\pm$0.15&34.2$\pm$4.7&180&0.71&6.9&13.2\\
g1\_3693&189.27757&62.25355&21.69$\pm$0.30&7.8$\pm$3.6&70&1.00&1.5&3.8\\
g1\_3694&189.28492&62.25408&20.29$\pm$0.09&50.3$\pm$4.9&222&0.98&10.5&25.8\\
\enddata
\tablecomments{(1) Identification (2) RA (J2000) (3) DEC (J2000) (4) J-magnitude (Vega) (5) Line flux (10$^{-17} erg s^{-1} cm^{-2}$ (6) Restframe equivalent width (\AA) (7) Extinction in magnitudes (8) SFR not corrected for extinction (M$_{\odot}$ yr$^{-1}$) (9) SFR corrected for extinction (M$_{\odot}$ yr$^{-1}$).}
\label{tab_objects}
\end{deluxetable}

\setcounter{table}{2}
\begin{deluxetable}{ccccrrcrr}
\tabletypesize{\scriptsize}
\tablecaption{Observed sample}
\tablehead{
\colhead{ID} & \colhead{$\alpha$ (J2000)} & \colhead{$\delta$ (J2000)} & \colhead{J}& \colhead{f$_{H_{\alpha}}$} & \colhead{EW(H$\alpha$)} & \colhead{A(H$\alpha$)} & \colhead{SFR$_{obs}$} & \colhead{SFR$_{cor}$} \\
\colhead{(1)} & \colhead{(2)} & \colhead{(3)} & \colhead{(4)}& \colhead{(5)} & \colhead{(6)} & \colhead{(7)} & \colhead{(8)}& \colhead{(9)}
}
\startdata

g1\_3756&189.36488&62.24973&21.58$\pm$0.22&14.7$\pm$5.2&261&1.63&3.1&14.1\\
g1\_3779&189.37307&62.24834&21.18$\pm$0.16&16.1$\pm$6.1&94&0.31&3.2&4.3\\
g1\_4171&189.02571&62.32623&22.01$\pm$0.22&3.5$\pm$2.5&40&1.28&0.7&2.2\\
g1\_4268&189.27285&62.33080&20.67$\pm$0.22&36.1$\pm$5.9&278&2.19&7.8&59.0\\
g1\_4284&189.27174&62.33083&20.41$\pm$0.19&9.8$\pm$3.7&80&0.08&1.9&2.0\\
g1\_4398&189.33542&62.32533&21.85$\pm$0.24&11.8$\pm$5.0&219&2.23&2.5&19.1\\
g1\_4423&189.39726&62.32167&19.42$\pm$0.06&36.7$\pm$7.8&42&2.21&6.9&52.8\\
g1\_4558&189.35632&62.31392&19.22$\pm$0.05&51.9$\pm$8.4&51&2.22&9.9&76.7\\
g1\_4806&189.39811&62.30147&18.99$\pm$0.05&32.3$\pm$10.6&23&0.94&5.9&14.0\\
g1\_4832&189.32014&62.30673&20.18$\pm$0.12&32.9$\pm$5.1&179&2.26&6.6&53.3\\
g1\_4908&189.42394&62.29461&21.93$\pm$0.21&26.4$\pm$6.7&272&2.63&6.3&70.8\\
g1\_5183&189.34373&62.28091&20.86$\pm$0.15&16.6$\pm$4.7&177&1.04&3.3&8.7\\
g1\_5226&189.17678&62.27920&21.12$\pm$0.19&9.7$\pm$2.8&64&1.95&1.9&11.2\\
g1\_5276&189.33590&62.27489&20.48$\pm$0.13&27.6$\pm$7.5&77&2.05&5.4&35.9\\
\enddata
\tablecomments{(1) Identification (2) RA (J2000) (3) DEC (J2000) (4) J-magnitude (Vega) (5) Line flux (10$^{-17} erg s^{-1} cm^{-2}$ (6) Restframe equivalent width (\AA) (7) Extinction in magnitudes (8) SFR not corrected for extinction (M$_{\odot}$ yr$^{-1}$) (9) SFR corrected for extinction (M$_{\odot}$ yr$^{-1}$).}
\label{tab_objects}
\end{deluxetable}

\end{document}